\documentclass[sigconf]{acmart}

\usepackage{tikz}
\usetikzlibrary{arrows}
\usepackage{amsmath}
\usepackage{todonotes}
\presetkeys{todonotes}{inline}{}
\usepackage{algorithm}
\usepackage[noend]{algpseudocode}
\usepackage{multirow}
\usepackage{graphicx}
\graphicspath{{figures/}}

\usepackage[subtle, lists=tight]{savetrees}

\usepackage{xcolor}
\definecolor{azure}{rgb}{0.54, 0.17, 0.89}

\usepackage{xspace}

\usepackage{graphicx} 
\usepackage{subcaption}
\usepackage{tabularx}
\usepackage{xspace}
\usepackage{listings}
\usepackage{verbatim}
\usepackage{booktabs}
\usepackage{colortbl}
\usepackage{nicefrac}
\usepackage{siunitx}
\usepackage{amsmath}
\usepackage{amsthm}
\usepackage{url}
\usepackage{microtype}

\usepackage{color}
\usepackage{url}
\definecolor{darkred}{rgb}{0.7,0,0}
\definecolor{darkgreen}{rgb}{0,0.5,0}

\newcommand{\bitcoin}{Bitcoin\xspace}
\newcommand{\bng}{Bitcoin-NG\xspace}
\newcommand{\prism}{Prism\xspace}
\newcommand{\newprism}{Prism++\xspace}
\newcommand{\algorand}{Algorand\xspace}
\newcommand{\ethereum}{Ethereum\xspace}

\newcommand{\indicator}{{\bf 1}}

\newcommand{\ma}[1] {{\textcolor{blue}{MA: #1}}}

\theoremstyle{definition}
\newtheorem{defn}{Definition}
\theoremstyle{remark}
\newtheorem{remark}{Remark}

\begin{document}

\title{Practical Low Latency Proof of Work Consensus}

\author{Lei Yang}
\affiliation{%
  \institution{MIT CSAIL}
  \country{}
}
\email{leiy@csail.mit.edu}

\author{Xuechao Wang}
\affiliation{%
  \institution{UIUC}
  \country{}
}
\email{xuechao2@illinois.edu}

\author{Vivek Bagaria}
\affiliation{%
  \institution{Matician}
  \country{}
}
\email{vivekee047@gmail.com}

\author{Gerui Wang}
\affiliation{%
  \institution{Beijing Academy of Blockchain and Edge Computing}
  \country{}
}
\email{wanggerui@baec.org.cn}

\author{Mohammad Alizadeh}
\affiliation{%
  \institution{MIT CSAIL}
  \country{}
}
\email{alizadeh@csail.mit.edu}

\author{David Tse}
\affiliation{%
  \institution{Stanford University}
  \country{}
}
\email{dntse@stanford.edu}

\author{Giulia Fanti}
\affiliation{%
  \institution{CMU}
  \country{}
}
\email{gfanti@andrew.cmu.edu}

\author{Pramod Viswanath}
\affiliation{%
  \institution{Princeton University}
  \country{}
}
\email{pramodv@princeton.edu}
\begin{abstract}
\bitcoin is the first fully-decentralized permissionless blockchain protocol to achieve a high level of security, but at the expense of poor throughput and latency. Scaling the performance of \bitcoin has a been a major recent direction of research. One successful direction of work has involved replacing proof of work (PoW) by proof of stake (PoS). 
Proposals  to scale the performance in the PoW setting itself have focused mostly on {\em parallelizing} the mining process,  scaling throughput; the few proposals to improve latency have either sacrificed throughput or the latency guarantees involve large constants rendering it practically useless.  Our first contribution is to design a new PoW blockchain \newprism that has provably low latency and high throughput; the design retains the parallel-chain approach espoused in \prism but invents a new confirmation rule to infer the permanency of a block by combining information across the parallel chains. We show   security at the level of \bitcoin with very small confirmation latency (a small constant factor of block interarrival time). A key aspect to scaling the performance is to use a large number of parallel chains, which puts significant strain on the system.  Our second contribution is the design and evaluation of a practical system to efficiently manage the memory, computation, and I/O imperatives of a large number of parallel chains. Our implementation of \newprism achieves a throughput of over $80,000$ transactions per second and confirmation latency of tens of seconds on networks of up to $900$ EC2 Virtual Machines.
\end{abstract}

{\def\addcontentsline#1#2#3{}\maketitle}

\section{Introduction}
In 2008, Satoshi Nakamoto invented \bitcoin and the concept of {\em blockchains}~\cite{bitcoin}. Since then, blockchains have attracted considerable interest for their applications in cross-border payments~\cite{hileman2017global,kazan2015value}, digital contracts~\cite{cong2019blockchain,wust2018you,pilkington201611} and more. At the heart of \bitcoin and many other blockchain projects is the {\em Nakamoto longest chain protocol}. It enables an open (permissionless) network of nodes to reach consensus on an ordered log of transactions. 

A distinguishing feature Nakamoto's protocol is the use of proof-of-work (PoW) consensus. A node gets the right to propose a block of transactions by solving a cryptographic puzzle, a process called ``mining''. Each mined block refers to a previous block, extending a chain, and an honest miner always tries to extend the longest chain. A block is ``confirmed'' when it is embedded $k$-deep in the longest chain, where $k$ is a parameter controlling the probability of reversal, i.e. the probability that another chain could ever overtake the longest chain. 
The protocol is secure against Byzantine adversaries controlling no more than 50\% of the network’s compute power. However, achieving this high level of security comes at severe cost to throughput and latency, since it requires the block mining time to be long to reduce forking and the confirmation depth $k$ to be large to make the reversal probability small.  \bitcoin, for example, supports 3--7 transactions per second and can take hours to confirm a transaction with a high level of reliability~\cite{bitcoin}.  

Improving the throughput of PoW consensus has received significant attention in recent years. Proposals range from simply increasing the block size~\cite{bitcoinsize} to restructuring the protocol to scale throughput arbitrarily on top of a slowly growing (and hence secure) longest chain~\cite{bitcoin-ng, ohie, fruitchains}. Improving {\em latency}, however, is much more difficult.  No existing protocol can achieve low latency (e.g., a small multiple of the network delay) while maintaining the security of longest chain protocol. A recent attempt, Prism~\cite{prism-theory}, took an important step: sortition blocks into proposer blocks and voter blocks, and the voter blocks are further sortitioned into multiple parallel ``voter chains’’ to reduce confirmation latency. The idea is to confirm a proposer block using voter blocks from  many chains, each mined independently using the longest chain protocol. With many votes, one can confirm much faster, because although each such vote is not very deep in its respective longest chain (hence not reliable), overall high reliability can still be achieved because there are many independent voting chains.

While the idea of multiple voter chains is elegant, the Prism protocol’s latency is far from ideal. Confirming blocks requires a confirmation rule to compute the reversal probability as a function of the state of the voter chains (analogous to the standard k-deep rule for the longest chain protocol). Prism provides such a rule and proves that it guarantees latency that is a constant multiple of the network delay. But the constant is huge and crucially depends on the time horizon of the blockchain’s operation, making the bound useless in practice. Our analysis shows that Prism's confirmation rule implies a latency bound of several hours depending on the adversarial mining power assumed. To ensure security, there is no choice but to wait until the confirmation rule guarantees the desired level of reliability; we show that in practice this leads to a high latency in \prism. Moreover, the reliability of the confirmation rule depends on using a large number of voter chains (tens of thousands), which is difficult if not impossible to realize in real systems. 


The second challenge is that parallel voting chain protocols are much more complex to implement than the longest chain protocol. Nodes must process hundreds to thousands of blocks per second in real time, putting significant strain on system. The high block rate and intricate state updates impose a particularly high I/O requirements on databases and persistent storage. Thus it is unclear to what extent the theoretical performance properties can translate to real-world performance. 

\begin{figure} 
    \centering
    \includegraphics{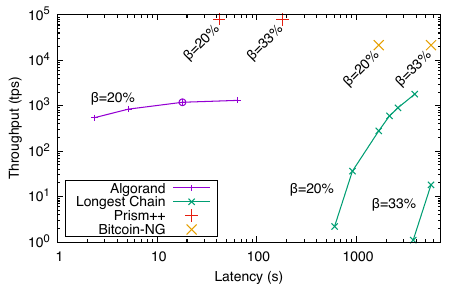}
    \caption{\small Throughput and confirmation latency of \newprism, \algorand, \bng, and the longest chain protocol on the same testbed. Note that the axes are on log scales. For \algorand and the longest chain protocol, parameters are tuned to span an optimized tradeoff between throughput and latency at a given security level (details in \S\ref{sec:eval}). 
     For \bng and \newprism, throughput and latency are decoupled so one can simultaneously optimize both at one operating point for a given security level. However, the throughput of Bitcoin-NG drops to that of the longest chain protocol under attack, while that of \newprism remains high. More details in \S\ref{sec:related} and \S\ref{sec:eval-performance}.
    }
    \label{fig:compare}
\end{figure}

In this paper, we present \newprism, the first practical PoW consensus protocol to achieve low latency, high throughput, and Bitcoin-level security. \newprism solves the two challenges above by providing the following contributions:
\begin{itemize}
  \item The first confirmation rule for parallel voting chains that has provably low latency and Bitcoin-level security. This rule is designed to enable explicit identification of the worst-case adversarial attack, which allows us to sharply characterize the confirmation error probability in a general, continuous-time model over an infinite time horizon of operation. Identifying worst-case attacks on consensus protocols is challenging, and our approach of co-designing a confirmation rule along with a worst-case attack is novel, to the best of our knowledge. This methodology also allows us to sharply characterize the worst-case latency of confirmation, which we show to be a small constant of the block interarrival time (e.g., for adversarial hash power $\beta$ of 20\%, the latency bound is roughly a factor of 12 of the block interarrival time). This is elaborated in   \S\ref{sec:new_rule}.

  \item An end-to-end system implementation and evaluation of \newprism, showing that it can achieve both low latency and high throughput (near theoretical predictions) in practice. Our implementation, written in 10,000 lines of Rust, supports pay-to-public-key UTXO transactions (similar to \bitcoin) and includes several crucial performance optimizations, such as asynchronous ledger updates and a scoreboarding technique to enable multi-threaded transaction execution. We evaluate it on a testbed of up to 900 EC2 machines connected via an emulated wide-area network. Fig.~\ref{fig:compare} shows the main results: \newprism achieves 30$\times$ lower latency than the longest chain protocol, and delivers a high throughput of 80,000 tps.
  \end{itemize}

\noindent {\bf Network and  Security Model}. We adopt the now-standard  $\Delta-$ 
synchronous network model for longest chain protocol analysis  \cite{ren2019analysis,dembo2020everything};  here the adversary controls the delivery of messages but the end-to-end network delay between any two nodes is upper bounded by $\Delta$. Also, following standard convention \cite{bitcoin,ren2019analysis,dembo2020everything,gazi2020tight}, we model the block mining process as Poisson with rate $\lambda$ proportional to the hash power of the miner. 
The adversary is assumed to control a fraction $\beta$ of the total hash power.

\smallskip
\noindent {\bf Organization}. 
The rest of the paper is organized as follows. In \S\ref{sec:related} we overview different  approaches to scaling  blockchain performance. In \S\ref{sec:scaling} we provide a short background on the \prism protocol and explain its limitations. In \S\ref{sec:new_rule}, we propose our novel confirmation rule for parallel voting chain, along with formal guarantees on security and low worst-case latency. \S\ref{sec:design} discusses the practical challenges when turning the algorithm into a working system, as well as the details of the client implementation with an interface enabling pay-to-public-key transactions. 

\section{Related Work }
\label{sec:related}
There are broadly three different approaches to scale the performance of blockchains. First, \textit{on-chain scaling} aims to design consensus protocols with inherently high throughput and low  latency. Protocols such as Bitcoin-NG~\cite{bitcoin-ng}, GHOST~\cite{ghost}, Algorand~\cite{algorand}, OHIE~\cite{ohie} are examples of this approach. 
Second, in \textit{off-chain scaling}, users establish cryptographically-locked agreements called ``payment channels''~\cite{payment-channel} and send most of the transactions off-chain on those channels. 
Lightning~\cite{lightning} and Eltoo~\cite{decker2018eltoo} are examples of this approach. 
Third, \textit{sharding} approaches conceptually maintain multiple ``slow'' blockchains that achieve high performance in aggregate. Omniledger~\cite{kokoris2018omniledger}, Ethereum 2.0~\cite{buterin2016ethereum}, and Monoxide~\cite{monoxide} are examples of this approach. 
These three approaches are orthogonal and can be combined to aggregate their individual performance gains. 

Since \newprism is an on-chain scaling solution, we compare it with other on-chain solutions. 
We explicitly exclude protocols with different trust and security assumptions, like   Tendermint~\cite{kwon2014tendermint},  HotStuff~\cite{yin2019hotstuff}, HoneyBadgerBFT~\cite{miller2016honey}, SBFT~\cite{gueta1804sbft}, Stellar~\cite{stellarsystem}, and Ripple\cite{cachin2017blockchain}, which require clients to pre-configure a set of validators who are responsible for executing the consensus protocol. 
These protocols target ``permissioned'' settings, and they generally scale to significantly fewer validator nodes than the above-mentioned permisionless protocols.


Among protocols with similar security assumptions to ours,  Bitcoin-NG~\cite{bitcoin-ng}  mines blocks  at a low rate similar to the longest chain protocol. In addition, each block's miner continuously adds transactions to the ledger until the next block is mined. This utilizes the capacity of the network between the infrequent mining events, thereby improving throughput, but latency remains the same as that of the longest-chain protocol. Furthermore,  an adversary that adaptively corrupts miners can reduce its throughput to that of the longest chain protocol by censoring the addition of transactions~\cite{parallel}. \newprism adopts the idea of decoupling the addition of transactions from the election into the main chain but avoids this adaptive attack. We compare to \bng in \S\ref{sec:eval-performance}.

DAG-based solutions like GHOST~\cite{ghost}, Inclusive~\cite{inclusive}, and Conflux~\cite{conflux} were designed to operate at high mining rates, and their blocks form a directed acyclic graph (DAG). 
However, these protocols were later shown to be insecure because they don't guarantee liveness, i.e. the ledger stops growing under certain balancing attacks \cite{ghost_attack}. Spectre\cite{spectre} and Phantom \cite{phantom} protocols were built along the ideas in GHOST and Inclusive to defend against the balancing attack, however, they don't provide any formal guarantees. Also, Spectre doesn't give a total ordering and
Phantom has a liveness attack \cite{conflux}.
To the best of our knowledge, the GHOST, Inclusive, Spectre and Phantom protocols have no publicly available implementation, and hence we were not able to compare these protocols with \newprism in our performance evaluation. 

\newprism maintains the same blockchain structure as \prism: a DAG with many parallel chains, and a clear separation of blocks into different types with different functionalities (Figure \ref{fig:prism}). 
OHIE~\cite{ohie} and Parallel Chains~\cite{parallel} build on these lessons by running many slow, secure longest chains in parallel, which gives high aggregate throughput at the same latency as the longest-chain protocol. To our knowledge, Parallel Chains has not been implemented. 
In OHIE's latest implementation~\cite{ohiecode}, transactions are placeholder signed messages that do not depend on prior transactions; because of this, clients do not need to maintain the unspent transaction output (UTXO) state of the blockchain. 
This makes comparison with \cite{ohiecode} difficult because in practice, we observed that transaction processing (including UTXO state management) is a major computational bottleneck for high-throughput systems (see \S\ref{sec:eval}).

Algorand~\cite{algorand} takes a different approach by adopting a proof of stake consensus protocol and tuning various parameters to maximize the performance. We compare to Algorand in Figure~\ref{fig:compare} and in \S\ref{sec:eval-performance}. Importantly, none of the above protocols simultaneously achieve both high throughput and low latency. 
Their reported throughputs are all lower than \newprism's, and their latencies are all higher than \newprism's, except for Algorand which has a lower latency.



\section{Background}
\label{sec:scaling}

Several recent blockchain proposals have succeeded in scaling the throughput of the longest chain, such as Bitcoin-NG \cite{bitcoin-ng}, GHOST \cite{ghost}, and Fruitchains \cite{fruitchains}. However, how to scale the latency has been a major challenge in blockchain research for years. \newprism borrows an idea from \prism~\cite{prism-theory}, a recent attempt to scale PoW consensus, by systematically decoupling the functions of the blocks in the longest chain protocol. Specifically, \newprism retains the mining and blockchain structure of \prism. In this section, we provide an overview of this idea, and explain why \prism does not deliver low latency in practice.

\subsection{Prism and Decoupling the Longest Chain}
\label{sec:prism}

\begin{figure}
\begin{center}
\includegraphics[width=0.45\textwidth]{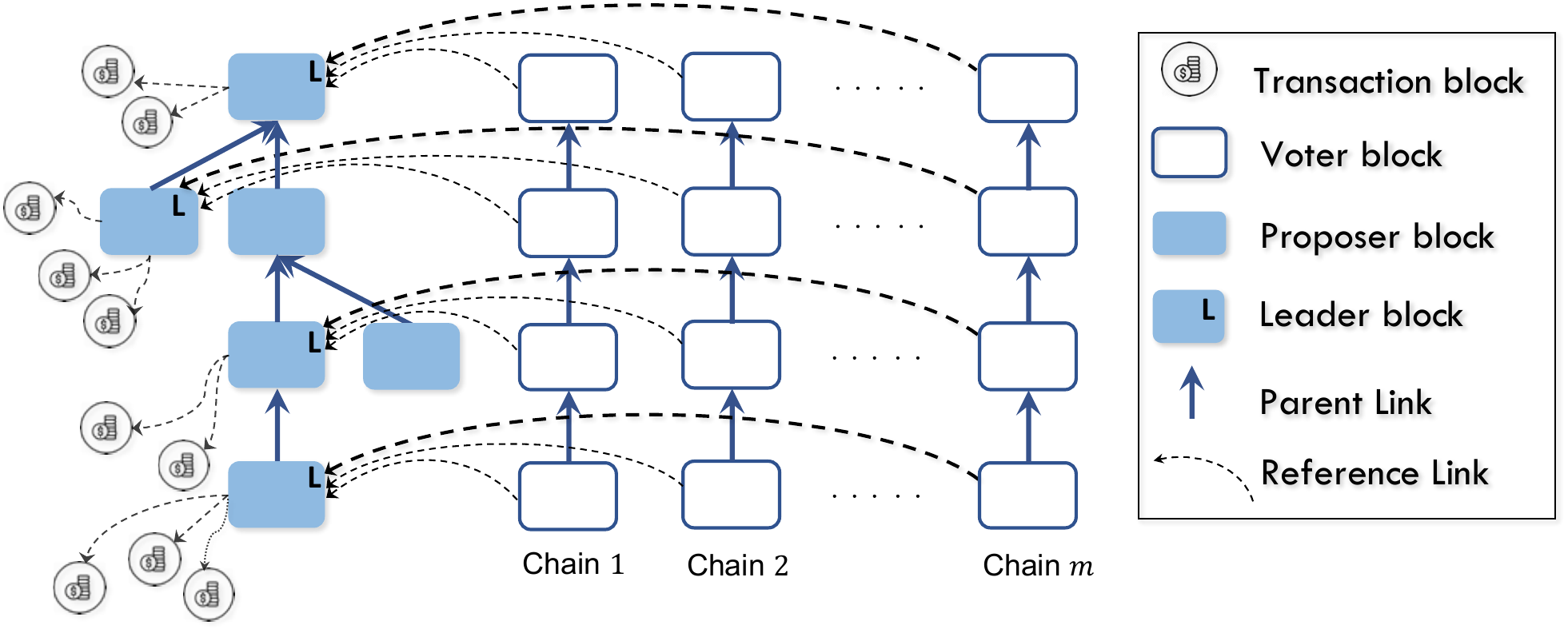}
\end{center}
\caption{\small Factorizing the blocks into three types of blocks: proposer blocks, transaction blocks and voter blocks.}
\label{fig:prism}
\end{figure}

The selection of a main chain in a blockchain protocol can be viewed as electing a leader block among all the blocks at each level of the blocktree. In this light, the blocks in the longest chain protocol can be viewed as serving three distinct roles: they stand for election to be leaders;  they add transactions to the main chain; they vote for ancestor blocks through parent link relationships. The latency and throughput limitations of the longest chain protocol are due to the {\em coupling} of the roles carried by the blocks. \prism~\cite{prism-theory} proposes to remove these limitations by factorizing the blocks into {\em three types} of blocks: {\em proposer blocks}, {\em transaction blocks}, and {\em voter blocks} (Figure \ref{fig:prism}). Each block mined by a miner is randomly sortitioned into one of the three types of blocks, and if it is a voter block, it will be further  sortitioned into one of several voter chains. (Mining is described in more detail below.)

The \textit{proposer} blocktree anchors the blockchain. 
Each proposer block contains a list of reference links to \textit{transaction} blocks which contain transactions, as well as a single reference to a parent proposer block.
Honest nodes mine proposer blocks on the longest chain in the proposer tree, but the chain structure does \textit{not} determine the final confirmed sequence of proposer blocks.
We define the \emph{level} of a proposer block as its distance from the genesis proposer block, and the \emph{height} of the proposer tree as the maximum level that contains any proposer blocks.
To determine the ordering of proposer blocks (and thus transaction blocks and transactions), we elect one \textit{leader} proposer block from each level. The sequence of leader blocks up to the height of the proposer tree is called the  \textit{leader sequence}, and is determined by the \emph{voter} chains. 


There are $m$ voter chains, where $m \gg 1$ is a fixed parameter chosen by the system designer. For example, we choose $m=1000$ in our experiments. New blocks are mined on each voter chain according to the longest chain rule. A voter block votes for a proposer block by including its hash, and the votes are cast such that each voter chain votes for one and only one proposer block at each level of the proposer tree. We consider a vote \textit{valid} if it is cast by a voter block on a longest voter chain. For each level in the proposer tree, we define the proposer block that receives the most valid votes as the \textit{leader block} (ties are broken by the hash of the proposer blocks). The sequence of the leader blocks decides a unique ordering of the transaction blocks, which forms the final ledger (ledger formation is explained in more detail below).

\smallskip \noindent
\textbf{Mining}.\label{sec:mining} In Prism, miners should not be able to choose \emph{a priori} which type of block they are mining; this is essential for the security of the scheme, since otherwise the adversary could concentrate its mining power on a subset of block trees and overpower the honest nodes.
\emph{Cryptographic sortition} is used to ensure that miners cannot choose which type of block they mine.
Nodes combine the contents of the transaction block, the proposer block, and the $m$ voter blocks (one for each voter chain) they are mining into a \emph{superblock} and perform proof-of-work mining on it.
We divide the range of a valid proof of work into $m+2$ sections, each corresponding to one type of block.
The concrete type \textbf{T} of a superblock is then decided by looking at which section its proof-of-work falls into.
Here, the proof-of-work value essentially serves as a random oracle
 that dictates the type of a mined block.
However, there is still one problem: once we know the concrete type \textbf{T} of a superblock, we want it to contain only the content of type \textbf{T} but not other types, otherwise we will waste bandwidth. As a solution, miners summarize the contents of the $m+2$ candidate blocks using a Merkle tree, and perform mining on the Merkle root instead of the plain hash of the combined content. When the concrete type \textbf{T} of the superblock is decided, the miner keeps only the content of type \textbf{T} and inserts a Merkle proof to show that it had committed to this particular content when the block was mined.

\smallskip \noindent
\textbf{Ledger Formation}.\label{sec:confirmation} In Prism, transaction blocks are mined at a high rate.
A key consequence is that transaction blocks mined concurrently may contain redundant or conflicting transactions.
If Prism were to discard blocks that contain inconsistent transactions, it would waste bandwidth because transactions that \emph{are} consistent are also discarded.
To prevent this, Prism separates the process of confirming blocks and forming a ledger.
The key observation is that given a confirmed leader sequence of proposer blocks, the ordering of transaction blocks is uniquely determined. This gives us a total ordering of transactions since the inception of the protocol, and enables us to determine the validity of each individual transaction by executing them following this ordering. For example, among transactions that try to spend the same coin, only the first is considered valid.

\smallskip \noindent
\textbf{Summary}. At a high level, Prism achieves high throughput by decoupling it from security: transactions are carried by separate transaction blocks and hence throughput can be increased by simply increasing the mining rate of transaction blocks, while the security stays the same because the mining rate of proposer blocks remains low. The throughput is only limited by the computing or communication bandwidth $C$ of each node, thus enabling a $100\%$ utilization.
For latency, the key idea is that even though it takes a long time to achieve low reversal probability of \textit{individual} votes (same as in \bitcoin), the collective opinion on leader blocks can converge much faster because there are many parallel voter chains. Next, we will discuss some nuances of this idea, and explain why Prism does not deliver low latency in practice.

\subsection{Limitations of Prism}
\label{sec:limitation_Prism}


\noindent {\bf \prism's confirmation rule~\cite{prism-theory}}. The confirmation rule in a blockchain defines when nodes can consider a block to be part of the immutable ledger. For example, the longest chain protocol famously uses the $k$-deep confirmation rule. In \prism, the ledger is determined by the proposer leader blocks, so the confirmation rule needs to decide when the leader block of a proposer level is immutable.

The immutability of a leader block depends on the immutability of the votes, which in turn depends on how deep they appear in the respective voter chains, or, their \textit{depths}. The key challenge is to consider the depths of all $m$ votes for a proposer level, such that we can confirm a leader block quickly, and the leader block will not change once confirmed (except for negligible probability). 

 For each proposer block at a level, \prism calculates a {\em confidence interval} for the total number of votes it will \textit{eventually} receive. When the confidence interval of any single proposer block is above that of all other blocks on the same level, it confirms the block. Bagaria et al.~\cite{prism-theory} prove a security theorem for this rule in a round-by-round synchronous model. We now explain the limitations of this confirmation rule regarding security and confirmation latency. 

\noindent {\bf Limitations of security guarantee}. In \cite{prism-theory}, the confidence interval of the number of votes a proposer block receives depends on a finite bound on the lifespan of the blockchain in rounds ($r_{\rm max}$). As $r_{max}$ increases, the probability of a successful attack increases in this analysis (via a union bound). Specifically, the exact expression of error probability is $r_{\rm max}^2 e^{-\frac{(1-2\beta)m}{16 \log{m}}}$. To compensate for the effect of $r_{\rm max}$, \cite{prism-theory} increases the number of voter chains ($m$) to keep the confirmation error probability low.
Table \ref{table:num_voter} shows the number of voter chains $m$ needed for a low error probability ($\epsilon = 10^{-9}$), which is impractically large and grows with the expected life span of the blockchain. 

\begin{table}
	\centering
	\caption{Relationship between the number of voter chains $m$ and the life span of the blockchain according to the security theorem in \cite{prism-theory}. We target a confirmation error probability of $\epsilon = 10^{-9}$. \cite{prism-theory} adopts a round-by-round model and we set the round duration to 1 second.}
	\begin{tabular}{ c || c | c | c | c | c | c } 
	 \hline
	  & \multicolumn{3}{c|}{$\beta = 0.2$} & \multicolumn{3}{c}{$\beta = 0.33$} \\
	 \hline
	 $r_{\rm max}$ (year)  & 1  &10  &100  & 1  &10  &100  \\ 
	 \hline
	 $m$ & 14075 & 15388  & 16716 &  26479 & 28939 & 31474 \\
	 \hline
	\end{tabular}
	\label{table:num_voter}
\end{table}

\noindent {\bf Limitation of the latency guarantee}. Although the confirmation rule in \cite{prism-theory} guarantees to confirm transactions within a constant factor of the network latency, the constant introduced in the analysis is large and leads to poor latency in practice.
To be specific, Theorem 4.8 in \cite{prism-theory} shows that the confirmation latency of an honest transaction (a transaction without public double spending) is upper bounded by $c_1(\beta) = \frac{5400(1-\beta)}{(1-2\beta)^3 \log{\frac{1-\beta}{\beta}}} \log{\frac{50}{1-2\beta}}$ rounds when $m$ is very large. Table \ref{table:old_latency} lists this bound for several different $\beta$, which is too large to provide any practical guarantee.

\begin{table}
	\centering
	\caption{Upper bound on the confirmation latency guaranteed by \cite{prism-theory}. \cite{prism-theory} adopts a round-by-round model and we set the round duration to be 1 second. The poor latency bounds are due to the large constants introduced in the analysis.}
	\begin{tabular}{ c || c | c | c } 
	 \hline
     $\beta$ &0.1  & 0.2 & 0.33 \\
	 \hline
	 Latency upper bound & 4.96h & 17.72h & 180h \\ 
	 \hline
	\end{tabular}
	\label{table:old_latency}
\end{table}

\medskip
\noindent {\bf \newprism's approach}. At the time of the analysis in \cite{prism-theory}, the state-of-the-art proof technique for longest chain  protocols were still rudimentary  \cite{backbone,pss16}, relying on finite execution time horizons. The proof techniques for the longest chain protocol, and corresponding security models and guarantees, have significantly advanced. Latest techniques can model continuous-time block arrivals with infinite time horizon  \cite{ren2019analysis,li2020continuous}, and provide sharp characterizations of the relation between network delay $\Delta$ and adversary fraction $\beta$ for secure blockchain operation \cite{dembo2020everything,gazi2020tight}. 
However, it is still unclear how to bring these techniques (or the ones in \cite{cryptoeprint:2020:675} on parallel ledger combining) to bear on the idea of \prism and its security analysis, given the multiple parallel voting chains coupled via a single block sortition procedure. We take a different approach in this paper: identifying the worst case attack and directly calculating the associated confirmation probability, {\em without resorting to a union bound}. We elaborate on our approach in the next section.
\section{A New Confirmation Rule for \prism}
\label{sec:new_rule}

In this section, we propose the confirmation rule of \newprism, a simple confirmation rule for Prism-style parallel voting that is secure with high probability and guarantees fast confirmation.

\noindent {\bf Intuition}. The high-level idea of the confirmation rule is that we can confirm a proposer block $B$ at a certain level $\ell$ when no other proposer block at level $\ell$ can ever receive more votes than $B$ in the future. Suppose that an $h$-fraction of voter chains have \textit{stabilized} with high probability, i.e., their votes will not change in the future. Further, suppose that $V$  voter chains vote for some other proposer block instead of $B$ at level $\ell$. 
Then, with high probability, $B$ will eventually receive at least $h m - V$ votes while any other proposer block at level $\ell$ could receive at most $V + m(1-h)$ votes. As a result, we can confirm $B$ if $h m-V>V+m(1-h)$, i.e., $h>\frac{V}{m}+\frac{1}{2}$. Note that $V$ is observable from the blockchain, so the only task now is to estimate $h$.

The key observation is that $h$ is a monotonically increasing function of time. This is because each voter chain follows the longest chain rule, where a prefix of the longest chain will stabilize as the chain grows longer (which happens as time passes). Although time is not observable in a PoW blockchain, it can be estimated from the growth of the voter chains. The large number of voter chains in \prism makes this estimation accurate even for short time periods because of the law of large numbers. 

\noindent {\bf Confirmation rule}.  Let $\tau_\ell$ be the time when a proposer block at level $\ell$ is mined or received by an honest node for the first time. According to the $\Delta$-synchronous model, at $\tau_\ell + \Delta$, all honest nodes will start voting for a proposer block at level $\ell$. 

Let $B_\ell(t)$ be the proposer block at level $\ell$ that receives the most votes at time $\tau_\ell + \Delta + t$, and let $\bar V_\ell(t)$ be the number of votes received by proposer blocks at level $\ell$ other than $B_\ell(t)$.
Let $d_i(t)$ be the depth of the voter block on the $i$-th voter chain which votes for a proposer block of level $\ell$ at time $\tau_\ell +\Delta + t$. If the $i$-th voter chain has not voted for any block on level $\ell$ at time $\tau_\ell +\Delta + t$, then $d_i(t)=0$.

For a single voter chain, let $q_t(\lambda_h,\lambda_a,\Delta,\pi)$ be the success probability of a certain attack $\pi$ to reverse the voter block that votes for a proposer block at level $\ell$ after time $\tau_\ell+\Delta+t$. Here, $\lambda_h = (1-\beta)\lambda$ is the honest mining rate, $\lambda_a = \beta \lambda$ is the adversarial mining rate, and $\Delta$ is the upper bound of the network delay.
Let $\pi^*(\lambda, \Delta)$ be the worst-case attack in terms of maximizing the success probability $q_t$.
Then, we define the {\em effective vote} for a proposer block at level $\ell$ from a voter chain at time $t + \tau_\ell+\Delta$ to be the fraction 
\begin{equation*}
    h^\Delta(t) := 1 - q_t(\lambda_h,\lambda_a,\Delta,\pi^*). 
\end{equation*}
We will derive an expression for $h^\Delta(t)$ later in this section. Now with all these definitions, we can propose the \newprism confirmation rule.

\begin{defn}[\newprism Confirmation Rule]
\label{def:rule}
We confirm $B_\ell(T)$ at time $\tau_\ell + \Delta + T$ if 
$$h^\Delta\left(\frac{1}{(1+\delta)m\lambda}\sum_{i=1}^{m} d_i(T)\right) \geq \frac{1}{m} \bar V_\ell(T) + \frac{1}{2} +\delta$$ 
for some small constant $\delta >0$ which controls the confirmation confidence. Note that $\tau_\ell + \Delta + T$ is an arbitrary time that an honest node chooses to observe the blockchain.
\end{defn}

\subsection{Deriving the New Confirmation Rule}

\noindent {\bf Outline}. Although the confirmation rule described above is simple, its security proof hinges on several building blocks. We start by focusing on the ideal case where the network delay is zero ($\Delta = 0$), all users are online (so that they observe when blocks appear), and only one public proposer block exists on each level. This allows us to focus on the voter chains.

We first revisit the security of the $k$-deep confirmation rule in the longest chain protocol, and introduce an alternative time-based confirmation rule (applicable for online users) on the single voter chain case. This time-based confirmation rule is readily extended to \prism-style parallel voting (with the proof for security similarly extensible) and achieves a much shorter confirmation latency compared to the longest chain protocol for the same reversal probability. Finally, we strengthen the time-based confirmation rule to handle the general setting: (a) offline users (making the confirmation rule only depend on the depth of the voter chains); (b) non-zero network delay $\Delta$; (c) multiple proposer blocks on the same level. This leads to the new \newprism confirmation rule as in Definition \ref{def:rule}.

{\bf \noindent \prism with one voter chain}. Before heading into the full Prism structure, we first take a look at the behavior of \prism with one voter chain ($m=1$), where the stabilized prefix of the single voter chain selects one leader proposer block at each level. The voter chain follows the Nakamoto longest chain protocol, which has a simple $k$-deep confirmation rule: confirm a block when it is buried $k$-blocks deep in the longest chain. This rule is remarkable in that it can be verified by nodes that join the protocol at any time, even much later than when the confirmed blocks first arrived. If we make the assumption that all nodes are {\em online} so that they observe when exactly a block arrives, then we can confirm a voter block (or equivalently the proposer block voted by it) based on time: a proposer block at level $\ell$ is confirmed $t$ seconds after the first block at level $\ell$ is received. Here, $t$ is a parameter to be chosen. We call this time-based confirmation rule the $t$-wait confirmation rule of this single voter chain scheme.

The reversal probability (the probability that a different proposer block at the same level will be selected in the future) of the $t$-wait confirmation rule is also the success probability of the worst-case attack  $q_t(\lambda_h,\lambda_a,0,\pi^*)$ by our definition. For the special case of $\Delta = 0$, a recent work~\cite{dembo2020everything} identified the worst-case attack on the $k$-deep rule: Theorem 5.1 of \cite{dembo2020everything} shows that the \textit{private attack} (with pre-mining proposed in \cite{sompolinsky2016bitcoin}) is the worst-case attack for every sample path. Following a similar logic, we prove that the same private attack is the worst case attack on the $t$-wait rule (see Appendix~\ref{app:proof}). 
To simplify the notation, we let $q_t^0(\lambda_h,\lambda_a) \triangleq q_t(\lambda_h,\lambda_a,0,\pi^*)$.

\begin{figure}
    \centering
    \includegraphics[width=0.35\textwidth]{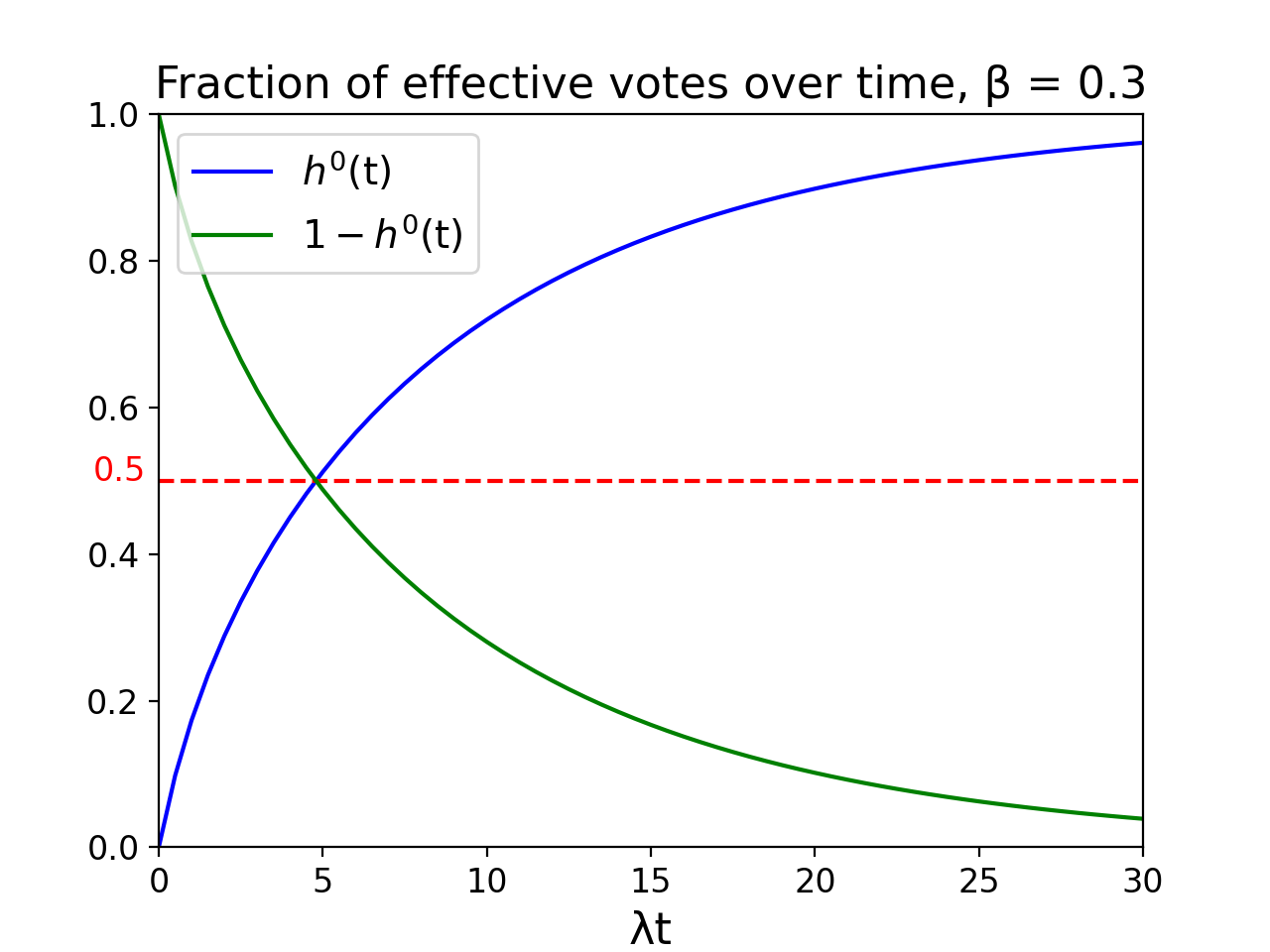}
    \caption{Fraction of effective votes on the single proposer block (blue) and a potential private block (green).}
    \label{fig:h0_t}
\end{figure}

In the pre-mining phase of the worst-case attack $\pi^*(\lambda, 0)$ (before any proposer block at level $\ell$ is received by all honest nodes, which, say, happens at time $\tau$), the adversary builds up a private chain with the maximum lead over the public voter chain, and the (random) length of such maximum lead at time $\tau$ is stochastically dominated by a Geometric distribution with parameter $p = 1-\frac{\lambda_a}{\lambda_h}$. After a proposer block is received by all honest nodes at time $\tau$, the honest nodes will start mining voter blocks to vote for that block, while the adversary starts a private attack from the maximum lead. The numbers of adversarial blocks and honest blocks mined in the time interval $(\tau, \tau+t)$ follow Poisson distributions with mean $\lambda_a t$ and $\lambda_h t$. If the pre-mined lead plus the newly-mined adversarial blocks is no less than the number of honest blocks at time $\tau+t$, then the attack succeeds; otherwise, from  simple random walk theory, the probability that the adversary will ever catch up from $z$ blocks behind is given by $\left(\frac{\lambda_a}{\lambda_h}\right)^z$ when $\lambda_a < \lambda_h$. 
Therefore, we have a closed-form expression for  $q_t^0(\lambda_h,\lambda_a)$: 
\begin{eqnarray*}
    &&~~~~~q_t^0(\lambda_h,\lambda_a) = 1 - \sum_{l = 0}^{\infty} \left(1 - \frac{\lambda_a}{\lambda_h}\right) \left( \frac{\lambda_a}{\lambda_h} \right)^l \cdot \\
    &&\left( \sum_{k = l}^{\infty} e^{-\lambda_h t} \frac{(\lambda_h t)^k}{k!} \cdot \sum_{n=0}^{k-l} e^{-\lambda_a t} \frac{(\lambda_a t)^n}{n!} \left (1-(\frac{\lambda_a}{\lambda_h})^{k-n-l} \right) \right).
\end{eqnarray*}
As a result, $h^0(t) = 1 - q_t^0((1-\beta)\lambda,\beta \lambda)$ is simply a function of $\lambda$ and $\beta$. 
Fig.~\ref{fig:h0_t} shows $h^0(t)$ for $\beta = 0.3$, where we see the exponential rise to unity. Recall that $h^0(t)$ is the probability that a proposer block is stabilized forever after duration $t$ since the block is mined. Indeed, this scheme achieves the same exponential security level as the longest chain protocol~\cite{ren2019analysis,backbone}, i.e., 
$$h^0(t) \geq 1 - A e^{-a\lambda t}$$
for large $t$, where positive constants $A$ and $a$ only depend on $\beta$. So, in order to achieve a very small error probability $\epsilon$ (we use $\epsilon = 10^{-9}$ in our experiments), the latency of this scheme $\tau_{single}$ should satisfy $h^0(\tau_{single}) \geq 1 - \epsilon$, i.e.,
\begin{equation*}
    \tau_{single} = O\left(\frac{1}{\lambda} \cdot \log{\frac{1}{\epsilon}}\right).
\end{equation*}
For example, we can compute that $\tau_{single} \approx 225/\lambda$ when $\beta = 0.3$ and $\epsilon = 10^{-9}$.

{\bf \noindent A time-based confirmation rule}. Returning to the Prism structure with many voter chains, we look for a similar time-based confirmation rule. Suppose an online node is aware of the arrival time $\tau_\ell$ of the only \textit{public} proposer block at level $\ell$ (\textit{private} proposer blocks could potentially exist), then when can the node confirm that proposer block? If the node had to wait long enough for every voter block that votes for this proposer block to stabilize, i.e., $h^\Delta \to 1$, then there would be no latency improvement over the longest chain protocol.

In fact, this is not necessary, because we only care about the sum of the votes. 
As long as $h^{\Delta}$, the probability of one voter block staying stable, is greater than 0.5, the adversary cannot reverse more than half of the votes except for probability exponentially small in $m$. This is because of the law of large numbers and the large deviation effect. 
So, when there is only one public proposer block at level $\ell$, we can simply confirm it when $h^{\Delta}(\cdot) > 1/2$. 
Although a vote from a particular voter chain may be reversed in the future with non-negligible probability (around $1/2$), the majority of the votes will stay stable when $m$ is large. The only challenge is that all online nodes may not agree on the proposer block arrival time $\tau_\ell$ due to network delay -- but this is not a problem when $\Delta = 0$. As a result, assuming $\Delta = 0$, we have the following time-based confirmation rule for Prism. 

\begin{defn}[CR1]
Assume nodes are aware of $\tau_\ell$, the arrival time of the only public proposer block at level $\ell$. Then, they confirm the proposer block at time $t_c$ such that $t_c = \tau_\ell + t_\delta^* $, where $h^0(t_\delta^*) = \frac{1}{2} + \delta$ for  some small constant $\delta > 0$ which controls the confirmation error probability.
\end{defn}

{\bf \noindent From time to depth}. Conventionally, we prefer a confirmation rule based on block depth rather than time, following the idea of the $k$-deep rule in the longest chain protocol. We can easily convert CR1 to a depth-based version: confirm the proposer block when more than $m\lambda t_\delta^*$ voter blocks are mined after $\tau_\ell$.
However, an offline node cannot execute this rule, because it does not know $\tau_\ell$. Instead, we can estimate the time by counting the depth of the voter blocks that vote for the public proposer block at level $\ell$, which is observable even for offline users. Therefore, we propose the following depth-based confirmation rule depending entirely on the {\em observable} blockchain.

\begin{defn}[CR2]
At time $\tau_\ell + t$, let $d_i(t)$ be the depth of the voter block on the $i$-th voter chain that votes for a proposer block of level $\ell$, or $d_i(t)=0$ if no such a voter block exists yet. Let $t_\delta^*$ satisfy $h^0(t_\delta^*) = \frac{1}{2} + \delta$ for $0 < \delta < \frac{1}{2}$ as in CR1. Then, at time $\tau_\ell + T$, if there is only one public proposer block on level $\ell$ and $\sum_{i=1}^{m} d_i(T) \geq (1+\delta)m\lambda t^*_\delta$, we can confirm this proposer block. Note that $\tau_\ell + T$ is an arbitrary time chosen by an honest node to observe the blockchain.
\end{defn}

\begin{remark}
Note that the conversion from time to block depth here is conservative as we assume each voter chain grows at the full rate $\lambda$ and the estimated time is an upper bound of the actual time elapsed. However, under the worst-case attack (private attack with pre-mining for $\Delta = 0$), each public voter chain grows at rate $(1-\beta) \lambda$. So a possibly better confirmation rule with lower latency would be to confirm the proposer block when $\sum_{i=1}^{m} d_i(T) \geq (1+\delta)m(1-\beta)\lambda t^*_\delta$. We conjecture this rule is also secure but defer a formal security proof as a future direction. 
\end{remark}


Note that under the normal path (with no adversarial action), each voter chain grows as a Poisson process with rate $\lambda$, so the confirmation latency of CR2 is concentrated around $t^*_\delta$ as $m$ is large. Later, we will show that for a fixed error probability $\epsilon$, we can choose very small slack $\delta$ by increasing $m$. Therefore, the confirmation latency is around $t^*_0$, where $h^0(t^*_0)=1/2$.
So the latency of \prism $\tau_{Prism}$ under the normal path would be
\begin{equation*}
    \tau_{Prism} = O\left(\frac{1}{\lambda}\right),
\end{equation*}
which is independent of the error probability $\epsilon$ when we choose $m$ to be large enough. For example, $\tau_{Prism} \approx 5/\lambda$ when $\beta = 0.3$. 

{\bf \noindent From $\Delta = 0$ to $\Delta > 0$}.  For a non-zero $\Delta$, characterizing the worst-case attack is challenging, but we are only interested in the regime where the mining rate of each voter chain is low, i.e., $\lambda \Delta \ll 1$. In Appendix \ref{app:proof_sketch}, we show that $q_t(\lambda_h,\lambda_a,\Delta, \pi^*(\lambda, \Delta)) =  q_t(\lambda_h,\lambda_a,\Delta, \pi^*(\lambda, 0)) + O(\lambda^2 \Delta^2)$. In words, for small $\lambda \Delta$, we can use the error probability under the private attack to approximate the error probability under the worst-case attack. 
To further simplify the calculation, we can approximate the honest chain growth using a Poisson process
with rate $\lambda_h' = \frac{\lambda_h}{1+\lambda_h \Delta}$ when $\lambda \Delta \ll 1$ as justified in Appendix \ref{app:latency}. 
So we have the approximation 
\begin{align*}
    h^\Delta(t) &= 1 - q_t(\lambda_h,\lambda_a,\Delta,\pi^*(\lambda, \Delta))  \\
    &\approx 1 - q_t^0(\lambda_h',\lambda_a).
\end{align*}

With this approximation, we can have a confirmation rule similar to CR2 for positive $\Delta$ when there is only one public proposer block. 

\begin{defn}[CR3]
At time $\tau_\ell +\Delta + t$, let $d_i(t)$ be the depth of the voter block on the $i$-th voter chain that votes for a proposer block of level $\ell$, or $d_i(t)=0$ if no such a voter block exists yet. Let $t_\delta^*$ satisfy $h^\Delta(t_\delta^*) = \frac{1}{2} + \delta$ for $0 < \delta < \frac{1}{2}$ as in CR1. At time $\tau_\ell + \Delta + T$, if there is only one public proposer block on level $\ell$ and $\sum_{i=1}^{m} d_i(T) \geq (1+\delta)m\lambda t^*_\delta$, then we confirm this proposer block. Note that $\tau_\ell + \Delta + T$ is an arbitrary time chosen by an honest node to observe the blockchain.
\end{defn}

{\bf \noindent Full \newprism confirmation rule}. So far, we have considered the  normal path where there is only one public proposer block when confirmation happens.
In practice, there could be multiple proposer blocks at the same level either due to a natural fork or an adversarial action. In a very extreme case (with low probability), the adversary may balance votes perfectly between two public proposer blocks, in which case there is no chance to confirm a proposer block any faster than the longest chain protocol. However, this is highly unlikely since we can require honest nodes to vote for the proposer block that receives the most votes -- in this case, the votes converge fast and such a balancing attack is unlikely to succeed.

We now complete the confirmation rule for Prism in the most general case where there could be multiple proposer blocks at any level $\ell$.
Let $B_\ell(t)$ be the proposer block at level $\ell$ that receives the most votes at time $\tau_\ell + \Delta + t$, and $\bar V_\ell(t)$ be the number of votes received by proposer blocks at level $\ell$ other than $B_\ell(t)$.
By our definitions, at time $\tau_\ell + \Delta +t$, about $m \cdot h^\Delta(t)$ voter chains are stabilized forever. So at least $m \cdot h^\Delta(t) - \bar V_\ell(t)$ voter chains that vote for $B_\ell(t)$ are stabilized. Since other proposer blocks can receive at most $\bar V_\ell(t) + m \cdot (1-h^\Delta(t))$ votes in the future, we can confirm $B_\ell(t)$ if $m \cdot h^\Delta(t) - \bar V_\ell(t) >\bar V_\ell(t) + m \cdot (1-h^\Delta(t))$, i.e., $h^\Delta(t)>\frac{1}{m} \bar V_\ell(t) + \frac{1}{2}$.

In this confirmation rule, $\bar V_\ell$ can be observed by counting the votes from each voter chain, but the time $t+\Delta$ passed after the appearance time $\tau_\ell$ is not observable for offline users. 
We estimate  time as in CR2 by counting  the  depth  of  the  voter  blocks  that  vote  for a proposer block at level $\ell$.
This leads us to  the rule for full \newprism confirmation rule, defined  in Definition \ref{def:rule}.

\subsection{Main Result}
Here, we state the core security property of this new confirmation rule for \prism blockchain structure. 

\begin{theorem}[Security of \newprism Confirmation Rule]
\label{thm:main}
If $B_\ell(T)$ is confirmed at time $\tau_\ell + \Delta + T$ such that the rule in Definition \ref{def:rule} is satisfied, i.e., $h^\Delta(\frac{1}{(1+\delta)m\lambda}\sum_{i=1}^{m} d_i(T)) \geq \frac{1}{m} \bar V_\ell(T) + \frac{1}{2} +\delta$,
then the number of votes received by $B_\ell(T)$ will be greater than the number of votes received by any other proposer block at any time $t>\tau_\ell + \Delta + T$, except for probability $e^{-\Omega(\delta^2 m)}$. $\tau_\ell + \Delta + T$ is an arbitrary time chosen by an honest node to observe the blockchain.
\end{theorem}

The proofs have been delegated to Appendices \S\ref{sec:delta=0} and \S\ref{app:proof_sketch}  due to space limitations. 

Our second main contribution is in identifying the worst attack to slow down the confirmation time, denoted as the ``slow-down attack" (defined formally in Appendix \S\ref{app:latency}). Such an identification along with the simplicity of the confirmation rule allow us to bound the worst case latency, as stated below.

\begin{theorem}[Latency of \newprism Confirmation Rule]
\label{thm:latency}
If $B_\ell$ is the only public proposer block at level $\ell$, then it will be confirmed with latency upper bounded by $O(\Delta)$, except for probability $e^{-\Omega(\delta^2 m)}$.
\end{theorem}

\begin{figure}
    \centering
    \includegraphics[width=0.35\textwidth]{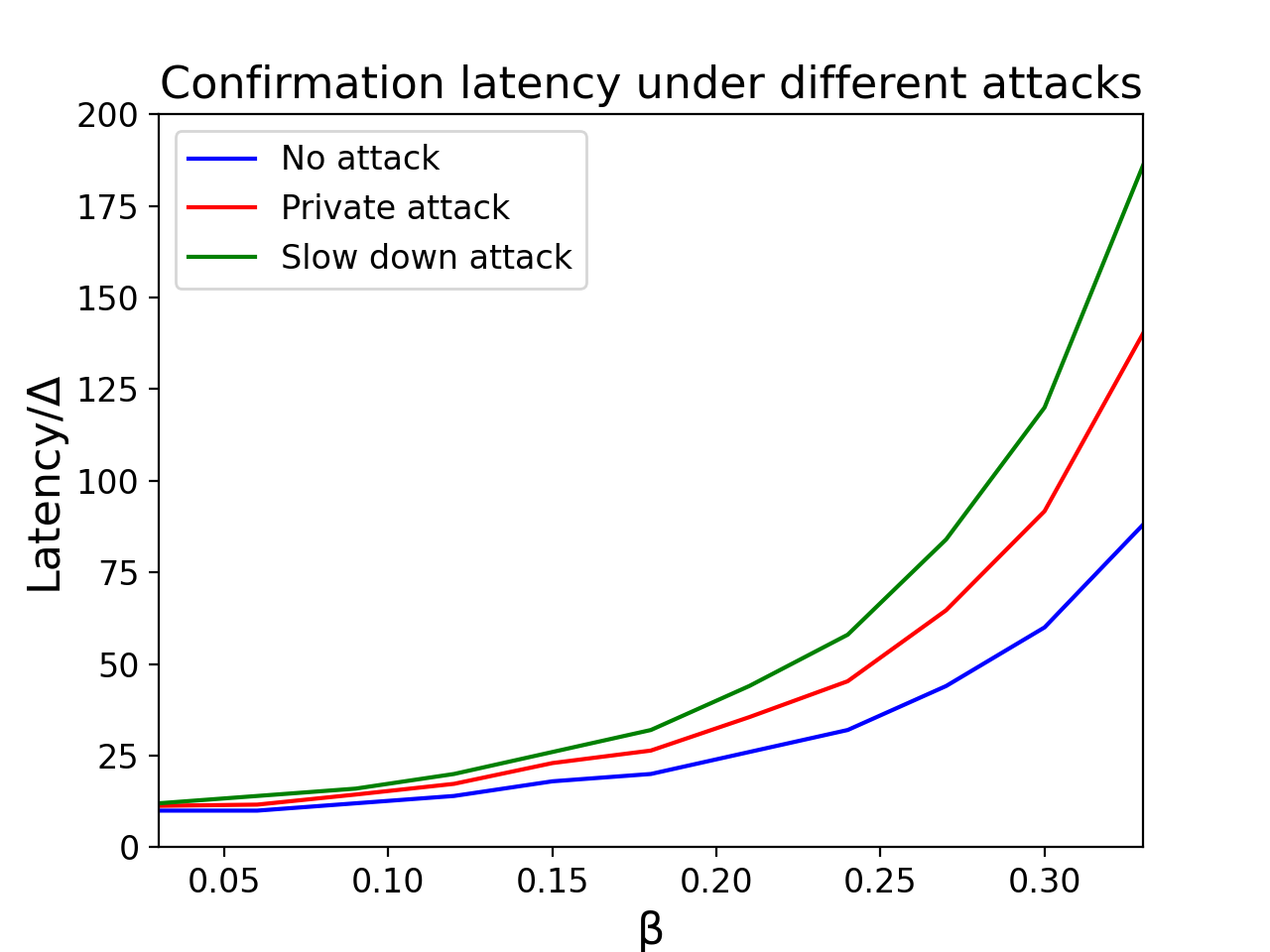}
    \caption{Latency under three attacks for different $\beta$. The latency is normalized over the upper bound on network latency  $\Delta$.}
    \label{fig:latency}
\end{figure}

The proof is deferred to Appendix \ref{app:latency}. The constants involved in Theorem~\ref{thm:latency} are all small and can be explicitly evaluated. A sample evaluation is illustrated in  Fig.~\ref{fig:latency} which plots the confirmation latency for different $\beta$  under three scenarios: no attack, private mining attack, and the worst case (slow-down) attack; here $\lambda \Delta = 0.1$ and $\delta = 0$. 

\section{System Implementation}
\label{sec:design}
An end-to-end cryptocurrency system based on the \prism structure is significantly more complex than one based on a single longest chain. 
This makes it challenging to design and implement a client software that can approach physical network limits without encountering other bottlenecks (e.g I/O). 
In this section, we describe the design and implementation of the \newprism client, the first client that achieves high-throughput and low-latency PoW consensus.

\label{sec:implementation-architecture}
\subsection{Architecture}

Our implementation is based on the \textit{unspent transaction output (UTXO)} model, similar to that used by \bitcoin. 
We support a simplified version of Bitcoin's scripting language, processing only pay-to-public-key (P2PK) transactions, similar to that implemented in Algorand~\cite{algorand, algorandcode}. We use Ed25519~\cite{ed25519} for cryptographic signatures and SHA-256~\cite{sha256} as the hashing algorithm.

\begin{figure}
    \centering
    \resizebox{0.8\columnwidth}{!}{\tikzstyle{ledger} = [draw, fill=blue!20, rectangle, 
    minimum height=4em, minimum width=6em, text centered, text width=5em]
\tikzstyle{blockchain} = [draw, fill=red!20, rectangle, minimum height=4em, minimum width=6em, text centered, text width=5em]
\tikzstyle{miner} = [draw, fill=green!20, rectangle, minimum height=4em, minimum width=6em, text centered, text width=5em]

\tikzstyle{database} = [draw, fill=yellow!20, rectangle, rounded corners, text width=5em, minimum height=3em, minimum width=6em, text centered]

\begin{tikzpicture}[auto, node distance=2.8cm,>=latex']
    \node [database] (blockchaindb) {Block Structure Database};
    \node [blockchain, above=0.5cm of blockchaindb] (blockchain) {Block Structure Manager};
    \node [ledger, left of=blockchaindb] (ledger) {Ledger Manager};
    \node [miner, right of=blockchaindb] (miner) {Miner};
    \node [database, left of=ledger] (utxodb) {UTXO Database};
    \node [database, right of=blockchain] (mempool) {Memory Pool};
    \node [database, left of=blockchain] (blockdb) {Block Database};
    \node [above=0.5cm of blockchain] (peers) {Peers};
    \node [right of=peers] (newtx) {New Transactions};
    \draw [<->] (blockchaindb) -- node[name=a] {} (blockchain);
    \draw [->] (blockchaindb) -- node[name=b] {} (miner);
    \draw [<->] (blockchaindb) -- node[name=c] {} (ledger);
    \draw [<-] (blockchain) -- node[name=d] {} (miner);
    \draw [<->] (miner) -- node[name=e]{} (mempool);
    \draw [->] (blockchain) -- node[name=f]{} (mempool);
    \draw [<->] (blockchain) -- node[name=g]{} (blockdb);
    \draw [<->] (ledger) -- node[name=h]{} (utxodb);
    \draw [<-] (ledger) -- node[name=i]{} (blockdb);
    \draw [<->] (peers) -- node[name=j]{} (blockchain);
    \draw [->] (newtx) -- node[name=k]{} (mempool);

\end{tikzpicture}}
    \caption{\small Architecture of our \newprism client implementation.}
    \label{fig:system-architecture}
\end{figure}

The system architecture is illustrated in Figure~\ref{fig:system-architecture}. Functionally it can be divided into the following three modules:

\begin{enumerate}
    \item \textit{Block Structure Manager}, which maintains the clients' view of the blockchain, and communicates with peers to exchange new blocks.
    \item \textit{Ledger Manager}, which updates the ledger based on the latest blockchain state, executes transactions, and maintains the UTXO set.
    \item \textit{Miner}, which assembles new blocks.
\end{enumerate}

\noindent 
The ultimate goal of the \newprism client is to maintain up-to-date information of the blockchain and the ledger. They are stored in the following four data structures:

\begin{enumerate}
    \item \textit{Block Structure Database}, residing in persistent storage, stores the graph structure of the blockchain (i.e., the voter blocktrees, the proposer blocktree, and the transactions blocks referenced) as well as the confirmed ordering of proposer and transaction blocks.
    \item \textit{Block Database}, residing in persistent storage, stores every block a client has downloaded so far. 
    \item \textit{UTXO Database}, residing in persistent storage, stores the UTXO set.
    \item \textit{Memory Pool}, residing in memory, stores the set of transactions that have not been mined into any block.
\end{enumerate}


At the core of the \textbf{Block Structure Manager} are an \textit{event loop} which sends and receives network messages to/from peers, and a \textit{worker thread pool} which handles those messages. When a new block arrives, the worker thread first checks its proof of work and sortition, according to the rules specified in \S\ref{sec:mining}, and stores the new block in the Block Database (checking proof of work at the earliest opportunity reduces the risk of DDoS attacks).  
It then proceeds to relay the block to peers that have not received it. Next, the worker thread checks whether all blocks referred by the new block, e.g. its parent, are already present in the database. If not, it buffers the block in an in-memory queue and defers further processing until all the referred block have been received. Finally, the worker validates the block (e.g., verifying transaction signatures), and inserts the block into the Block Structure Database. If the block is a transaction block, the Block Structure Manager also checks the Memory Pool against the transactions in this new block and removes any duplicates or conflicts from the Memory Pool.

The \textbf{Ledger Manager} is a two-stage pipeline and runs asynchronously with respect to the Block Structure Manager. Its first stage, the \textit{transaction sequencer}, runs in a loop to continuously poll the Block Structure Database and try to confirm new transactions. It starts by counting the votes cast on each proposer block. To avoid doing wasteful work, it caches the vote counts and the tips of the voter chains, and on each invocation, it only scans through the new voter blocks. Then, it tries to confirm a leader for each level in the proposer block tree where new votes are cast, according to the rules specified in \S\ref{sec:scaling} Definition \ref{def:rule}.  If a leader is selected, it queries the Block Database to retrieve the transaction blocks confirmed by the new leader, and assembles a list of confirmed transactions. The list is passed on to the second stage of the pipeline, the \textit{ledger sanitizer}. This stage maintains a pool of worker threads that executes the confirmed transactions in parallel. Specifically, a worker thread queries the UTXO Database to confirm that all inputs of the transaction are unspent; their owners match the signatures of the transaction;
and the input fund is sufficient. If execution succeeds, the outputs of the transaction are inserted into the UTXO Database, and the inputs are removed.

The \textbf{Miner} module assembles new blocks according to the mining procedure described in \S\ref{sec:mining}. It is implemented as a busy-spinning loop. At the start of each round, it polls the Block Structure Database and the Memory Pool to update the block it is mining. It also implements the spam mitigation mechanism described in \S\ref{sec:spamming-design}.
Like other academic implementations of PoW systems \cite{ohie,conflux}, our miner does not actually compute hashes for the proof of work, and instead simulates mining by waiting for an exponentially-distributed random delay. Solving the PoW puzzle in our experiments would waste energy for no reason, and in practice, PoW will happen primarily on dedicated hardware, e.g., application-specific integrated circuits (ASICs). So the cost of mining will not contribute to the computational bottlenecks of the consensus protocol.

The three databases residing in the persistent storage are all built on RocksDB~\cite{rocksdb}, a high-performance key-value storage engine. We tuned the following RocksDB parameters to optimize its performance: replacing B-trees with hash tables as the index; adding bloom filters; adding a 512 MB LRU cache; and increasing the size of the write buffer to 32 MB to sustain temporary large writes. 

\subsection{Important Optimizations}

\label{sec:implementation-highlights}

We have implemented a \newprism client in about 10,000 lines of Rust code.
The key challenge to implementing the \newprism client is to handle high throughput. The client must process blocks at a rate of hundreds of blocks per second, or a throughput of hundreds of Mbps, and confirm transactions at a high rate, exceeding 80,000 tps in our implementation. To handle the high throughput, our implementation exploits opportunities for parallelism in the protocol and carefully manages race conditions to achieve high concurrency. We now discuss several key performance optimizations. 


\noindent\textbf{Asynchronous Ledger Updates.}
In traditional blockchains like \bitcoin, blocks are mined at a low rate and clients update the ledger every time they receive a new block.  However, in the \newprism protocol, blocks are mined at a very high rate and a only a small fraction of these blocks\,---\,those that change the proposer block leader sequence\,---\,lead to changes in the ledger. Therefore, trying to update the ledger synchronously for each new block is wasteful and can introduce CPU bottleneck. 


Fortunately, \newprism does not require synchronous ledger updates to process blocks. Since \newprism allows conflicting or duplicate transactions to appear in the ledger and performs sanitization later (\S\ref{sec:confirmation}), the client need not update the ledger for each new block. Therefore, in our implementation, the Ledger Manager runs asynchronously with respect to the Block Structure Manager and periodically updates the ledger. 

Most blockchain protocols (e.g.,  \bitcoin, \algorand, and \bng) require that miners validate a block against the current ledger prior to mining it, and therefore cannot benefit from asynchronous ledger updates. For example, in \bitcoin's current specification, when a miner mines a block $B$, it implicitly certifies a ledger $L$ formed by tracing the blockchain from the genesis block to block $B$. 
A \bitcoin client receiving \textit{B} must therefore verify $B$ against the ledger $L$, and hence must update the ledger synchronously for each block. 
In principle, \bitcoin clients could perform \emph{post hoc} sanitization like \newprism; however, due to long block times relative to transaction verification, doing so would not improve performance.

\noindent\textbf{Parallel Transaction Execution.}
\label{sec:scoreboarding}
Executing a transaction involves multiple reads and writes to the UTXO Database to (1) verify the validity of the input coins, (2) delete the input coins, and (3) insert the output coins. If handled sequentially, transaction execution can quickly become the bottleneck of the whole system. Our implementation therefore uses a pool of threads in the Ledger Manager to execute transactions in parallel. Despite the  parallelism, the UTXO database is the bottleneck for the entire system (\S\ref{sec:eval-resource}).

However, naively executing all transactions in parallel is problematic, because semantically the transactions in the ledger form an order, and must be executed strictly in this order to get to the correct final state, i.e., UTXO set. For example, suppose  transactions \textit{T} and \textit{T'} both use UTXO \textit{u} as input, and \textit{T} appears first in the ledger. In this case, \textit{T'} should fail, since it tries to reuse \textit{u} when it has already been spent by \textit{T}. If we execute \textit{T} and \textit{T'} in parallel, race condition could happen where the inputs of \textit{T'} are checked before \textit{T} deletes \textit{u} from the UTXO set, allowing \textit{T'} to execute.

To solve this problem, we borrow the \textit{scoreboarding}~\cite{scoreboarding} technique long used in processor design. A CPU employing this method schedules multiple instructions to be executed out-of-order, as long as doing so will not cause conflicts such as writing to the same register. Transactions and CPU instructions are alike, in the sense that they both need to be executed in the correct order to produce correct results, only that transactions read and write UTXOs while CPU instructions read and write CPU registers. 
In the Ledger Manager, a batch of transactions first pass through a controller thread before being dispatched to one of the idle workers in the thread pool for execution. The controller keeps track of the inputs and outputs of the transactions in the batch on a ``scoreboard'' (an in-memory hash table). Before scheduling a new transaction for execution, it checks that none of its inputs or outputs are present on the scoreboard. In this way, all worker threads are able to execute in parallel without explicit synchronization. 

\noindent\textbf{No Transaction Broadcasting.}
In most traditional blockchains, clients exchange pending transactions in their memory pools with peers. This incurs extra network usage, because each transaction will be broadcast twice: first as a pending transaction, and again as part of a block. At the throughput of \newprism, such overhead becomes even more significant.

Our implementation does not broadcast pending transactions, because it is \textit{unnecessary} in \newprism. 
In traditional blockchains like Bitcoin and Ethereum, the whole network mines a block every tens of seconds or even few minutes. 
Since we cannot predict who will mine the next block, exchanging pending transactions is necessary, so that they get included in the next block regardless of who ends up mining it. In contrast, \newprism generates hundreds of transaction blocks every second. This elevated block rate means that any individual miner is likely to mine a transaction block in time comparable to the delay associated with broadcasting a transaction to the rest of the network (i.e., seconds). As a result, unlike other blockchain protocols, there is little benefit for a \newprism client to broadcast its transactions. Non-mining clients can submit their transactions to one or more miners for redundancy; however, miners do not need to relay those transactions to peers. 

\noindent\textbf{Spam Mitigation.}
\label{sec:spamming-design}
In \newprism, miners do not validate transactions before including them in blocks.
This introduces the possibility of spamming, where an adversary could generate a large number of conflicting transactions and send them to different nodes across the network.  
The nodes would then mine all of these transactions into blocks, causing miners and validators to waste storage and computational resources. We note that transaction fees alone cannot prevent such spamming. Since nodes only pay for transactions that make it to into the ledger, the adversary would not be charged for conflicting transactions that get removed during sanitization. 
Bitcoin is not susceptible to this attack because transactions are  validated  {\em prior to} mining. 
We propose a simple mechanism to mitigate spamming. Miners validate transactions with respect to their latest ledger state and other unconfirmed transactions, giving the adversary only a small window of network delay to spam the system. 
Miners then add a random delay prior to mining a transaction into blocks, thus increasing the chance that a miner detects a conflicting transaction in another transaction block, in which case it will drop the delayed spam transaction.
We evaluate the effectiveness of this method in \S\ref{sec:spamming-eval}.


\section{Evaluation}

\label{sec:eval}

Our evaluation answers the following questions:

\begin{itemize}
    \item By how much does the new confirmation rule in \newprism improve the confirmation latency over \prism? (\S\ref{sec:cfm-compare})
    \item What is the performance of \newprism in terms of transaction throughput and confirmation latency, and how does it compare with other protocols? (\S\ref{sec:eval-performance})
    \item How well does \newprism scale to a large numbers of users? (\S\ref{sec:eval-scale})
    \item How does \newprism perform with limited resource, and how efficient does it utilize resource? (\S\ref{sec:eval-resource})
    \item How does \newprism perform when under attack? (\S\ref{sec:eval-attack})
\end{itemize}


\noindent{\bf Schemes compared.} We compare the performance of \newprism with \algorand, \bng, and the longest chain protocol. We do not compare with \prism in most experiments, because its confirmation rule requires using more than 26,000 voter chains, which is impractical for real-world use. Instead, we run a targeted experiment to compare the latency of the \prism and \newprism confirmation rules under the same parameters (\S\ref{sec:cfm-compare}). For \prism, \bng, and the longest chain protocol, we modify and use our \newprism implementation. For \algorand, we use the official open-source implementation~\cite{algorandcode} written in Golang. Note that this implementation is different from the one evaluated in~\cite{algorand}. Therefore, we do not expect to reproduce the results in~\cite{algorand}.

\noindent{\bf Testbed.} We deploy our \newprism implementation on Amazon EC2's \texttt{c5d.4xlarge} instances with 16 CPU cores, 16 GB of RAM, 400 GB of NVMe SSD, and a 10 Gbps network interface. Each instance hosts one \newprism client. By default, we use 100 instances and connect them into a random 4-regular graph topology. To emulate a wide-area network, we introduce a one-way propagation delay of 120~ms on each link to match the typical delay between two distant cities~\cite{pingmeasurement}, and a rate limiter of 400~Mbps for ingress and egress traffic respectively on each instance. We also evaluate several other network topologies (up to 900 instances) and per-instance bandwidth limits.

To generate workloads for these experiments, we add a transaction generator in our testbed which continuously creates transactions at an adjustable rate. In our \newprism implementation, the main bottleneck is RocksDB and the I/O performance of the underlying SSD, which limits the throughput to about $90,000$ tps. 
We cap transaction generation rate to 80,000 tps to avoid hitting this bottleneck.


\noindent{\bf Performance tuning and security.} All protocols in the experiments have design parameters, and we tried our best to tune these parameters for performance and security. For \newprism, we calculate the optimal mining rate $f$ for proposer and voter blocks to achieve the best confirmation latency, given the adversarial ratio $\beta$ and desired confirmation confidence $\epsilon$. We cap the size of transaction blocks to be 40 KB, and set the mining rate for transaction blocks to support $80,000$ tps. Unless otherwise stated, we turn off the spam mitigation mechanism in \newprism (we evaluate its effectiveness in \S\ref{sec:eval-attack}).  To ensure security, we calculate the expected {\em forking rate} $\alpha$, i.e. fraction of blocks not on the main chain, given $f$ and the block propagation delay $\Delta$. We compare $\alpha$ against the forking rate actually measured in each experiment to ensure that the system has met the target security level. We follow the same process for \bng and the longest chain protocol. For \algorand, we adopt the default security parameters set in its production implementation. Then we hand-tune its latency parameters $\lambda$ and $\Lambda$. Specifically, we reduce $\lambda$ and $\Lambda$ until a round times out, and use the settings that yield the best confirmation latency. For \newprism, the longest chain protocol and \bng, we target a confirmation confidence $\epsilon$ in the order of $10^{-9}$. For \algorand, the blockchain halts with a probability in the order of $10^{-9}$.

\subsection{The New Confirmation Rule}
\label{sec:cfm-compare}

We have shown in \S\ref{sec:limitation_Prism} that the confirmation rule proposed in \prism~\cite{prism-theory} does not \textit{guarantee} low latency due to large constants. Here, we experimentally show that the rule does not \textit{achieve} optimal latency in practice by comparing it with the \newprism confirmation rule under the same parameters.

For both rules, we target a confirmation confidence of $\epsilon = 10^{-9}$. As explained in \S\ref{sec:limitation_Prism}, \prism's rule requires a large number of voter chains that depends on $\beta$, $\epsilon$, and the blockchain life span. To ensure a life span of at least 10 years under $\beta=0.3$ (the highest $\beta$ we explore in this experiment), \prism requires $m>23800$, so we set $m=23800$. We point out that this is impractically large for a real system, making the rule undeployable. In comparison, our new rule does not have such a limitation. \newprism may run for an infinite life span, and $m$ can be set independently from $\beta$ and $\epsilon$ (indeed, we use $m=1000$ for \newprism in all other experiments). Finally, we set the mining rate to 0.05 block/s per chain.

We compare the confirmation latency of \prism and \newprism under this setting. Results in Table \ref{table:real_latency} show that although the actual latency of the \prism confirmation rule is much lower than the upper bound (several hours, see \S\ref{sec:limitation_Prism}) given by \cite{prism-theory}, it is still 4--10$\times$ worse than \newprism. In comparison, the \newprism rule confirms blocks with almost-certain confidence ($\epsilon=10^{-9}$) when votes are only 3--9 blocks deep on average, achieving low latency in practice. 

\begin{table}[htb]
	\centering
	\caption{Actual confirmation latency of the \prism rule \cite{prism-theory} and the \newprism rule (\S\ref{sec:scaling}) under different $\beta$.}
	\begin{tabular}{ c || c | c | c } 
	 \hline
     $\beta$ &0.10  & 0.20 & 0.30 \\
	 \hline\hline
	 \newprism (s) & 55 & 82 & 184 \\
	 \hline
	 \prism (s) & 770   & 794 & 783 \\
	 \hline
	\end{tabular}
	\label{table:real_latency}
\end{table}

\subsection{Throughput and Latency}
\label{sec:eval-performance}

In this experiment, we measure the transaction throughput and confirmation latency of \newprism at different adversarial ratios $\beta$, and compare that with \algorand, \bng and the longest chain protocol. For \algorand, we use its default setting of security parameters, which targets $\beta=20\%$ (maximum possible is $\beta = 33\%$). 
 For \newprism, \bng and the longest chain protocol, we experiment with two adversarial ratios: $\beta=20\%$ and $\beta=33\%$. In both \algorand and the longest chain protocol, there is tradeoff between throughput and confirmation latency by choosing different block sizes. We explore this tradeoff and present it as a curve. For \algorand, we try block sizes between 300 KB to 32 MB. For the longest chain protocol, we try block sizes between 1.7 KB  to  33.6 MB. All four protocols are deployed on the same hardware and network topology as described above. We run each experiment for a minimum of 10 minutes and report the average transaction throughput and latency. The results are shown in Figure~\ref{fig:compare}.

\noindent{\bf Throughput.}
As shown in Fig. \ref{fig:compare}, \newprism is able to maintain the same transaction throughput of around $80,000$ tps regardless of the $\beta$ chosen.
This is because \newprism decouples throughput from security by using transaction blocks. In this way, \newprism is able to maintain the mining rate for transaction blocks to sustain a constant throughput, while changing the mining rate for other types of blocks to achieve the desired $\beta$. \bng offers a similar decoupling by entitling the miner of the latest key block to frequently produce micro blocks containing transactions. \algorand and the longest chain protocol do not offer such decoupling, so the block size must by increased in order to achieve a higher throughput. In such case, the confirmation latency increases, as demonstrated by the tradeoff curves in Figure~\ref{fig:compare}, to accommodate for the higher block propagation delay induced by larger blocks. 
For \algorand, we observe its throughput increases marginally with block size, but does not exceed $1300$ tps.
The reason is that \algorand only commits one block every round. So at any moment, unlike \newprism, \algorand only has one block propagating in the network, causing low bandwidth utilization. For \bng, we observed a peak throughput of 21,530 tps. The reason is that, unlike \newprism, in \bng only a single node (the leader) commits transactions at a time. This results in the network becoming a bottleneck; once the throughput exceeds about 20,000 tps, we observed that the block propagation delay increases significantly for \bng (note also that \bng is susceptible to an adaptive attack that censors the chosen leader and can reduce throughput substantially~\cite{parallel}).

\noindent{\bf Is Consensus the Throughput Bottleneck?}
A blockchain client has two roles: (1) it participates in the consensus protocol (the \textit{Block Structure Manager} and the \textit{Miner} in our implementation); (2) it executes transactions confirmed by the consensus protocol and updates the ledger (the \textit{Ledger Manager} in our implementation). The throughput can be bottlenecked by either of these stages and therefore we ask: Is the throughput limited by the consensus protocol, or the ledger updates? To answer this question, we measure the maximal throughput when no consensus protocol is involved, i.e. we start one client of each protocol and test how fast each client can execute transactions and update the ledger. For our \newprism, \bng and longest chain client, the limit is around $90,000$ tps. For \algorand, the limit is around $4,800$ tps. From Fig.~\ref{fig:compare} we see that \bitcoin, \bng, and \algorand have throughput much lower than these limits, and thus are bottlenecked by the consensus protocols. However, in the case of \newprism, its throughput is very close to the limit, and hence it is bottlenecked by the ledger updates.

\noindent{\bf Confirmation Latency.} At $\beta = 20\%$, \newprism achieves a latency of $42$ seconds, and for similar security guarantees \algorand under its default parameters achieves latency of $18$ seconds. Compared to the longest chain protocol, \newprism uses multiple voter chains in parallel ($1000$ chains in our experiments) to provide security instead of relying on a single chain. So \newprism requires each vote to be less deep in order to provide the same security guarantee. As a result, \newprism achieves a substantially lower confirmation latency. For example, for $\beta = 20\%$, the confirmation latency for \newprism is 42 seconds, compared to 607 seconds at the lowest throughput point for the longest chain protocol. As we increase the block size for the longest chain protocol, its  confirmation latency increases to 3870 seconds at a throughput of 1800 tps. The gap between \newprism and the longest chain protocol increases for higher $\beta$. For example, for \newprism the confirmation latency increases from 42 seconds to 182 seconds as $\beta$ increases from $20\%$ to $33\%$. For the longest chain protocol, the same change in $\beta$ causes the latency to increase by more than 3000 seconds. \bng exhibits similar confirmation latency as the longest chain protocol for the same value of $\beta$, since it applies the same $k$-deep rule as the longest chain protocol for key blocks to confirm transactions, and key blocks must be mined slowly to avoid frequent leader changes. Finally, we point out that the theoretical prediction of latency in \S\ref{sec:scaling} accurately translates to the real-world measurement. 
For example, the theory predicts a latency of 150 seconds for  $\beta=0.33$, while our measurement shows a latency of 182s. The extra 32 seconds is caused by (1) a transaction needs to wait for a proposer block to refer it; (2) there is a small probability that the proposer chain forks, and thus momentarily causes higher latency.


\subsection{Scalability}

\label{sec:eval-scale}
\begin{table}
	\centering
	\caption{\small Performance of \newprism with different network topologies.}
	\resizebox{\columnwidth}{!}{
	\begin{tabular}{ c || c | c c c } 
	 \hline
	 Property & \#Nodes & 100 & 300 & 900 \\ 
	 \hline\hline
	 \multirow{4}{*}{Degree $=4$}   & Diameter & 5 & 7 & 9 \\
	                                & Throughput (tps) & $8.1\times 10^4$ & $8.1\times 10^4$ & $7.8\times 10^4$ \\
	                                & Latency (s) & 42 & 71 & 76 \\
	                                & Forking & 0.15 & 0.17 & 0.15 \\
	 \hline
	 \multirow{4}{*}{Diameter $=5$} & Degree & 4 & 6 & 8 \\
	                                & Throughput (tps) & $8.1\times 10^4$ & $8.2\times 10^4$ & $8.0\times 10^4$ \\ 
	                                & Latency (s) &42 & 43 & 42 \\
	                                & Forking & 0.15 & 0.15 & 0.15 \\
	 \hline
	\end{tabular}
	}
	\label{table:scale}
	\end{table}

In this experiment, we evaluate \newprism's ability to scale to a large number of users. For each client, we use the same network and hardware configuration as in other experiments, and target an adversarial ratio $\beta=20\%$. The results are shown in Table~\ref{table:scale}.

First, we increase the number of clients while keeping the topology a random 4-regular graph, i.e., each client always connects to four random peers. In this case, the network diameter grows as the topology becomes larger, causing the block propagation delay to increase and the confirmation latency to increase correspondingly. Note that the transaction throughput is not affected because in \newprism the mining rate for transaction blocks is decoupled from that of the other types of blocks. Then, we explore the case where clients connect to more peers as the topology grows larger, so that the diameter of the network stays the same. As shown in the results, both confirmation latency and throughput are constant as the number of clients increases from 100 to 900. 

In all cases, the forking rate stays stable and is under $0.17$, proving that the system is secure for $\beta=20\%$. This suggests that \newprism is able to scale to a large number of users as long as the underlying peer-to-peer network provides a reasonable block propagation delay.

\subsection{Resource Utilization}
\label{sec:eval-resource}

In this experiment, we evaluate the resource utilization of our \newprism implementation, and how it performs with limited network bandwidth and CPU resources. 


\begin{figure}
\centering
\begin{minipage}[t]{.48\columnwidth}
  \centering
  \includegraphics[width=.9\textwidth]{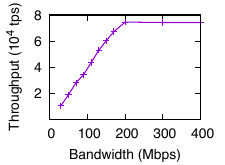}
    \caption{\small Performance of \newprism under different network bandwidth. }
    \label{fig:bw}
\end{minipage}\hfill%
\begin{minipage}[t]{.48\columnwidth}
  \centering
  \includegraphics[width=.9\textwidth]{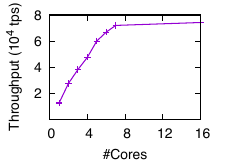}
    \caption{\small Performance of \newprism with different number of CPU cores at each client.}
    \label{fig:cpu}
\end{minipage}
\end{figure}

\begin{table}
\centering
\caption{\small Network bandwidth usage breakdown of \newprism measured on a 200 Mbps interface. Network Headroom is the unused bandwidth necessary for the block propagation delay to stay stable. Serialization overhead is wasted space when serializing in-memory objects for network transmission. Messaging stands for non-block messages.}
\resizebox{0.93\columnwidth}{!}{
\begin{tabular}{ c | c | c || r } 
 \hline
 \multicolumn{3}{c||}{Usage} & $\%$Bandwidth \\
 \hline\hline
 \multirow{5}{*}{Received} & \multirow{4}{*}{Deserialized} & Proposer Block & $0.05\%$ \\
 && Voter Block & $0.21\%$ \\
 && Transaction Block & $50.43\%$ \\ \cline{3-4}
 && Messaging &  $0.43\%$ \\ \cline{2-4} 
 & \multicolumn{2}{c||}{Serialization Overhead} & $25.80\%$ \\ \hline
 \multicolumn{3}{c||}{Network Headroom} & $23.08\%$ \\
 \hline
\end{tabular}
}

\label{table:bw-profiling}
\end{table}

\begin{figure*}
\centering
\begin{minipage}[t]{.30\textwidth}
  \centering
  \includegraphics[width=0.8\textwidth]{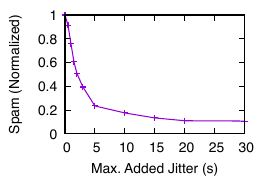}
    \caption{\small Effectiveness of random delay in defending against spam attack. Spam traffic amount is normalized to the case when no delay is added.}
    \label{fig:attack-spamming}
\end{minipage}\hfill%
\begin{minipage}[t]{.30\textwidth}
  \centering
  \includegraphics[width=0.8\textwidth]{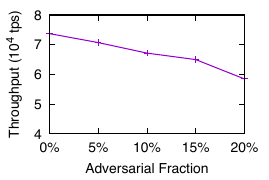}
    \caption{\small Performance of \newprism under  censorship attack.}
    \label{fig:attack-censor}
\end{minipage}\hfill%
\begin{minipage}[t]{.30\textwidth}
  \centering
  \includegraphics[width=0.8\textwidth]{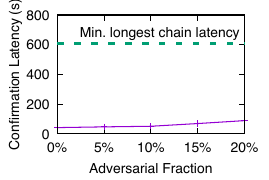}
    \caption{\small Performance of \newprism under balancing attack. We also mark the lowest confirmation latency possible for the longest chain protocol under the same security guarantee.}
    \label{fig:attack-balancing}
\end{minipage}
\end{figure*}

\noindent{\bf Network Bandwidth.} Figure~\ref{fig:bw} shows the throughput of \newprism as we throttle the bandwidth at each client. Results show that the throughput scales proportionally to the available bandwidth. The throughput stops to grow when the bandwidth is higher than 200 Mbps, because the transaction generation rate is capped at around 80,000 tps, which is near the bottleneck caused by RocksDB.

Table~\ref{table:bw-profiling} provides a breakdown of the bandwidth usage. Our implementation is able to process transaction data at a throughput about $50\%$ of the available bandwidth. (The in-memory size of a transaction is 168 bytes.) Further improvements could be made by using more efficient data serialization schemes and optimizing the underlying P2P network.

\noindent{\bf CPU.} Figure~\ref{fig:cpu} shows the throughput of \newprism as we change the number of CPU cores for each client. The throughput scales proportionally to the number of cores, and stops to grow after 7 cores because the transaction generation rate is capped. This shows that our implementation handles more than 10,000 tps per CPU core, and the parallelization techniques discussed in \S\ref{sec:implementation-highlights} are effective. 

Table~\ref{table:profiling} provides a breakdown of CPU usage across different components. More than half of CPU cycles are taken by RocksDB for which we only perform basic tuning. Less than $15\%$ are spent on overhead such as inter-thread communication, synchronization, etc. (categorized as ``Miscellaneous'' in the table). This suggests that our implementation uses CPU resources efficiently, and further improvements could be made primarily by optimizing the database. 

While we chose mid-end AWS EC2 instances for experiments, our results show that  \newprism  does not inherently require powerful machines or waste resources. For example, a laptop with 8 cores, 16 GB RAM, and 400 GB of NVMe-based SSD would cost under \$3,000 today and could easily run \newprism at 80,000 tps. 


\begin{table}
\centering
\caption{\small CPU usage breakdown of our \newprism implementation.}
\resizebox{0.65\columnwidth}{!}{
\begin{tabular}{ c | c || r } 
 \hline
 \multicolumn{2}{c||}{Operation} & $\%$CPU \\  
 \hline\hline
 \multirow{3}{*}{Ledger} & RocksDB Read/Write & $49.5\%$ \\ 
                         & (De)serialization & $3.1\%$\\ 
                         & Miscellaneous & $8.9\%$\\ \hline
 \multirow{5}{*}{Blockchain}    & Signature Check & $21.7\%$ \\ 
                         & (De)serialization & $3.8\%$ \\
                         & RocksDB Read/Write & $3.9\%$ \\
                         & Network I/O & $0.6\%$ \\ 
                         & Miscellaneous & $5.5\%$\\ \hline
  \multicolumn{2}{c||}{Block Assembly} & $1.5\%$ \\ \hline
 \multicolumn{2}{c||}{Transaction Generation} & $0.7\%$ \\ \hline
 \multicolumn{2}{c||}{Miscellaneous} & $0.8\%$ \\ \hline
\end{tabular}
}
\label{table:profiling}
\end{table}
\subsection{Performance Under Active Attack}
\label{sec:eval-attack}

In the following experiments, we evaluate how \newprism performs in the presence of active attacks. Specifically, we consider three  types of attacks: {\em spamming}, \textit{censorship}, and \textit{balancing} attacks. Spamming and censorship attacks aim to reduce network throughput, while balancing attacks aim to increase confirmation latency. 
In these experiments we configure \newprism to tolerate a maximum adversarial ratio $\beta=20\%$. 


\label{sec:spamming-eval}
\noindent{\bf Spamming Attack.} Recall that in a spamming attack, attackers send conflicting transactions to different nodes across the network. As described in \S\ref{sec:spamming-design}, miners can mitigate such attack by adding a random timing jitter to each transaction. In this experiment, we set up 100 miners as victims and connect them according to the same topology as in other experiments. Then for each miner we start a local process that generates a transaction every 100 ms. We synchronize those processes across the network so that each miner receives the same transaction at the same time, with a time synchronization error of several ms due to the Network Time Protocol. 
To defend against the attack, miners add a uniform random delay before including a transaction into the next transaction block. We let each attack to last for 50 seconds, and measure the fraction of spam transactions that end up in transaction blocks. 
Fig.~\ref{fig:attack-spamming} shows that adding a random jitter of at most 5 seconds can reduce the spam traffic by about 80\%. 
We point out that miners can extend this method by monitoring the reputation of clients by IP address and public key, and penalizing clients with high spam rate with longer jitter.


\noindent{\bf Censorship Attack.}
In a censorship attack, malicious clients mine and broadcast empty transaction blocks and proposer blocks. Censorship attack does not threaten the security of \newprism, but it reduces the system throughput because a fraction of blocks are now ``useless'' since they do not contain any data. As Figure~\ref{fig:attack-censor} shows, during a censorship attack, the transaction throughput reduces proportionally to the percentage of adversarial users, i.e. the system capacity degrades gracefully.

\label{sec:balancing}

\noindent{\bf Balancing Attack.}
In a balancing attack, attackers try to increase the confirmation latency of the system by waiting for the event when multiple proposer blocks appear on the same level and balancing the votes among them. Normally, when multiple proposer blocks appear on one level, every client votes for the proposer block with the most votes, so the system quickly converges with the vast majority of voter chains voting for one proposer block. During a balancing attack, however, the attacker votes on the proposer blocks with the second most votes to slow down such convergence, causing votes to be more evenly distributed among competing proposer blocks. In this case, clients need to wait for votes to grow deeper in order to confirm a proposer leader, resulting in longer confirmation latency. Figure~\ref{fig:attack-balancing} shows that the confirmation latency grows as the active adversarial fraction increases. But even when $20\%$ clients are malicious, the confirmation latency is still more than $6\times$ better than the longest chain protocol.

\section{Conclusion}
\label{sec:conclusion}

In this paper, we have presented \newprism, the first practical PoW consensus protocol to achieve low latency, high throughput, and Bitcoin-level security. \newprism retains the mining and the chain structure of \prism \cite{prism-theory}, and invents a new confirmation rule based on a novel methodology of co-designing the confirmation rule with an explicit identification of the worst-case attack. We proved that this confirmation rule is crucial in achieving low latency in practical settings. In the second part of the paper, we have converted the \newprism protocol to an efficient software system.  Our implementation supports over 80,000 transactions per second at a confirmation latency of tens of seconds with Bitcoin-level security. Our results validate the theoretical analysis of the new confirmation rule, and highlight the importance of optimizing transaction execution and the databases for high throughput. We also demonstrated experimentally that \newprism is robust to several active attacks, and showed that a simple random delay is effective at mitigating spamming. 

\section*{Acknowledgments}

We thank Ertem Nusret Tas for feedback and fruitful discussions.

\bibliographystyle{plain}
\bibliography{refs}{}

\begin{thebibliography}{10}

\bibitem{algorandcode}
algorand/go-algorand: Algorand's official implementation in go.
\newblock https://github.com/algorand/go-algorand.

\bibitem{bitcoinsize}
Block size limit controversy.

\bibitem{pingmeasurement}
Global ping statistics.
\newblock https://wondernetwork.com/pings/.

\bibitem{ohiecode}
ivicanikolicsg/ohie: Ohie - blockchain scaling.
\newblock https://github.com/ivicanikolicsg/OHIE.

\bibitem{rocksdb}
Rocksdb | a persistent key-value store.
\newblock https://rocksdb.org.

\bibitem{prism-theory}
Vivek Bagaria, Sreeram Kannan, David Tse, Giulia Fanti, and Pramod Viswanath.
\newblock Prism: Deconstructing the blockchain to approach physical limits.
\newblock In {\em Proceedings of the 2019 ACM SIGSAC Conference on Computer and
  Communications Security}, CCS '19, page 585–602, New York, NY, USA, 2019.
  Association for Computing Machinery.

\bibitem{ed25519}
Daniel~J. Bernstein, Niels Duif, Tanja Lange, Peter Schwabe, and Bo{-}Yin Yang.
\newblock High-speed high-security signatures.
\newblock {\em J. Cryptographic Engineering}, 2(2):77--89, 2012.

\bibitem{buterin2016ethereum}
Vitalik Buterin.
\newblock Ethereum 2.0 mauve paper.
\newblock In {\em Ethereum Developer Conference}, volume~2, 2016.

\bibitem{cachin2017blockchain}
Christian Cachin and Marko Vukoli{\'c}.
\newblock Blockchain consensus protocols in the wild.
\newblock {\em arXiv preprint arXiv:1707.01873}, 2017.

\bibitem{cong2019blockchain}
Lin~William Cong and Zhiguo He.
\newblock Blockchain disruption and smart contracts.
\newblock {\em The Review of Financial Studies}, 32(5):1754--1797, 2019.

\bibitem{decker2018eltoo}
Christian Decker, Rusty Russell, and Olaoluwa Osuntokun.
\newblock eltoo: A simple layer2 protocol for bitcoin.
\newblock {\em White paper: https://blockstream. com/eltoo. pdf}, 2018.

\bibitem{payment-channel}
Christian Decker and Roger Wattenhofer.
\newblock A fast and scalable payment network with bitcoin duplex micropayment
  channels.
\newblock In Andrzej Pelc and Alexander~A. Schwarzmann, editors, {\em
  Stabilization, Safety, and Security of Distributed Systems - 17th
  International Symposium, {SSS} 2015, Edmonton, AB, Canada, August 18-21,
  2015, Proceedings}, volume 9212 of {\em Lecture Notes in Computer Science},
  pages 3--18. Springer, 2015.

\bibitem{dembo2020everything}
Amir Dembo, Sreeram Kannan, Ertem~Nusret Tas, David Tse, Pramod Viswanath,
  Xuechao Wang, and Ofer Zeitouni.
\newblock Everything is a race and nakamoto always wins.
\newblock {\em arXiv preprint arXiv:2005.10484}, 2020.

\bibitem{bitcoin-ng}
Ittay Eyal, Adem~Efe Gencer, Emin~Gun Sirer, and Robbert~Van Renesse.
\newblock Bitcoin-ng: A scalable blockchain protocol.
\newblock In {\em 13th {USENIX} Symposium on Networked Systems Design and
  Implementation ({NSDI} 16)}, pages 45--59, Santa Clara, CA, March 2016.
  {USENIX} Association.

\bibitem{parallel}
Matthias Fitzi, Peter Ga{\v{z}}i, Aggelos Kiayias, and Alexander Russell.
\newblock Parallel chains: Improving throughput and latency of blockchain
  protocols via parallel composition.
\newblock Cryptology ePrint Archive, Report 1119, 2018.

\bibitem{cryptoeprint:2020:675}
Matthias Fitzi, Peter Gazi, Aggelos Kiayias, and Alexander Russell.
\newblock Ledger combiners for fast settlement.
\newblock Cryptology ePrint Archive, Report 2020/675, 2020.
\newblock \url{https://eprint.iacr.org/2020/675}.

\bibitem{backbone}
Juan Garay, Aggelos Kiayias, and Nikos Leonardos.
\newblock The bitcoin backbone protocol: Analysis and applications.
\newblock In {\em Annual International Conference on the Theory and
  Applications of Cryptographic Techniques}, pages 281--310. Springer, 2015.

\bibitem{gazi2020tight}
Peter Gaži, Aggelos Kiayias, and Alexander Russell.
\newblock Tight consistency bounds for bitcoin.
\newblock Cryptology ePrint Archive, Report 2020/661, 2020.
\newblock \url{https://eprint.iacr.org/2020/661}.

\bibitem{algorand}
Yossi Gilad, Rotem Hemo, Silvio Micali, Georgios Vlachos, and Nickolai
  Zeldovich.
\newblock Algorand: Scaling byzantine agreements for cryptocurrencies.
\newblock In {\em Proceedings of the 26th Symposium on Operating Systems
  Principles, Shanghai, China, October 28-31, 2017}, pages 51--68. {ACM}, 2017.

\bibitem{gueta1804sbft}
G.~{Golan Gueta}, I.~{Abraham}, S.~{Grossman}, D.~{Malkhi}, B.~{Pinkas},
  M.~{Reiter}, D.~{Seredinschi}, O.~{Tamir}, and A.~{Tomescu}.
\newblock Sbft: A scalable and decentralized trust infrastructure.
\newblock In {\em 2019 49th Annual IEEE/IFIP International Conference on
  Dependable Systems and Networks (DSN)}, pages 568--580, 2019.

\bibitem{hileman2017global}
Garrick Hileman and Michel Rauchs.
\newblock Global cryptocurrency benchmarking study.
\newblock {\em Cambridge Centre for Alternative Finance}, 33, 2017.

\bibitem{kazan2015value}
Erol Kazan, Chee-Wee Tan, and Eric~TK Lim.
\newblock Value creation in cryptocurrency networks: Towards a taxonomy of
  digital business models for bitcoin companies.
\newblock In {\em PACIS}, page~34, 2015.

\bibitem{kokoris2018omniledger}
Eleftherios Kokoris-Kogias, Philipp Jovanovic, Linus Gasser, Nicolas Gailly,
  Ewa Syta, and Bryan Ford.
\newblock Omniledger: A secure, scale-out, decentralized ledger via sharding.
\newblock In {\em 2018 IEEE Symposium on Security and Privacy (SP)}, pages
  583--598. IEEE, 2018.

\bibitem{kwon2014tendermint}
Jae Kwon.
\newblock Tendermint: Consensus without mining.
\newblock {\em Draft v. 0.6, fall}, 1:11, 2014.

\bibitem{inclusive}
Yoad Lewenberg, Yonatan Sompolinsky, and Aviv Zohar.
\newblock Inclusive block chain protocols.
\newblock In {\em International Conference on Financial Cryptography and Data
  Security}, pages 528--547. Springer, 2015.

\bibitem{conflux}
Chenxing Li, Peilun Li, Wei Xu, Fan Long, and Andrew Chi-chih Yao.
\newblock Scaling nakamoto consensus to thousands of transactions per second.
\newblock {\em arXiv preprint arXiv:1805.03870}, 2018.

\bibitem{li2020continuous}
Jing Li and Dongning Guo.
\newblock Continuous-time analysis of the bitcoin and prism backbone protocols.
\newblock {\em arXiv preprint arXiv:2001.05644}, 2020.

\bibitem{stellarsystem}
Marta Lokhava, Giuliano Losa, David Mazi\'{e}res, Graydon Hoare, Nicolas Barry,
  Eli Gafni, Jonathan Jove, Rafał Malinowsky, and Jed McCaleb.
\newblock Fast and secure global payments with stellar.
\newblock In {\em Proceedings of the 27th Symposium on Operating Systems
  Principles}. ACM, 2019.

\bibitem{miller2016honey}
Andrew Miller, Yu~Xia, Kyle Croman, Elaine Shi, and Dawn Song.
\newblock The honey badger of bft protocols.
\newblock In {\em Proceedings of the 2016 ACM SIGSAC Conference on Computer and
  Communications Security}, pages 31--42. ACM, 2016.

\bibitem{bitcoin}
Satoshi Nakamoto.
\newblock Bitcoin: A peer-to-peer electronic cash system.
\newblock 2008.

\bibitem{ghost_attack}
Christopher Natoli and Vincent Gramoli.
\newblock The balance attack against proof-of-work blockchains: The r3 testbed
  as an example.
\newblock {\em arXiv preprint arXiv:1612.09426}, 2016.

\bibitem{sha256}
US~NIST.
\newblock Descriptions of sha-256, sha-384 and sha-512, 2001.

\bibitem{pss16}
R~Pass, L~Seeman, and A~Shelat.
\newblock Analysis of the blockchain protocol in asynchronous networks.
\newblock In {\em Annual International Conference on the Theory and
  Applications of Cryptographic Techniques}, 2017.

\bibitem{fruitchains}
R.~Pass and E.~Shi.
\newblock Fruitchains: A fair blockchain.
\newblock In {\em Proceedings of the ACM Symposium on Principles of Distributed
  Computing}. ACM, 2017.

\bibitem{pilkington201611}
Marc Pilkington.
\newblock 11 blockchain technology: principles and applications.
\newblock {\em Research handbook on digital transformations}, 225, 2016.

\bibitem{lightning}
Joseph Poon and Thaddeus Dryja.
\newblock The bitcoin lightning network: Scalable off-chain instant payments,
  2016.

\bibitem{ren2019analysis}
Ling Ren.
\newblock Analysis of nakamoto consensus.
\newblock {\em IACR Cryptol. ePrint Arch.}, 2019:943, 2019.

\bibitem{spectre}
Y~Sompolinsky, Y~Lewenberg, and A~Zohar.
\newblock Spectre: A fast and scalable cryptocurrency protocol.
\newblock {\em IACR Cryptology ePrint Archive}, 2016:1159.

\bibitem{phantom}
Y~Sompolinsky and A~Zohar.
\newblock Phantom: A scalable blockdag protocol, 2018.

\bibitem{ghost}
Yonatan Sompolinsky and Aviv Zohar.
\newblock Secure high-rate transaction processing in bitcoin.
\newblock In {\em International Conference on Financial Cryptography and Data
  Security}, pages 507--527. Springer, 2015.

\bibitem{sompolinsky2016bitcoin}
Yonatan Sompolinsky and Aviv Zohar.
\newblock Bitcoin's security model revisited.
\newblock {\em arXiv preprint arXiv:1605.09193}, 2016.

\bibitem{scoreboarding}
James~E. Thornton.
\newblock Parallel operation in the control data 6600.
\newblock In {\em Proceedings of the October 27-29, 1964, Fall Joint Computer
  Conference, Part II: Very High Speed Computer Systems}, AFIPS '64 (Fall, part
  II), pages 33--40, New York, NY, USA, 1965. ACM.

\bibitem{monoxide}
Jiaping Wang and Hao Wang.
\newblock Monoxide: Scale out blockchains with asynchronous consensus zones.
\newblock In Jay~R. Lorch and Minlan Yu, editors, {\em 16th {USENIX} Symposium
  on Networked Systems Design and Implementation, {NSDI} 2019, Boston, MA,
  February 26-28, 2019.}, pages 95--112. {USENIX} Association, 2019.

\bibitem{wust2018you}
Karl W{\"u}st and Arthur Gervais.
\newblock Do you need a blockchain?
\newblock In {\em 2018 Crypto Valley Conference on Blockchain Technology
  (CVCBT)}, pages 45--54. IEEE, 2018.

\bibitem{yin2019hotstuff}
Maofan Yin, Dahlia Malkhi, Michael~K Reiter, Guy~Golan Gueta, and Ittai
  Abraham.
\newblock Hotstuff: Bft consensus with linearity and responsiveness.
\newblock In {\em Proceedings of the 2019 ACM Symposium on Principles of
  Distributed Computing}, pages 347--356, 2019.

\bibitem{ohie}
H.~Yu, I.~Nikolic, R.~Hou, and P.~Saxena.
\newblock Ohie: Blockchain scaling made simple.
\newblock In {\em 2020 IEEE Symposium on Security and Privacy (SP)}, pages
  90--105, Los Alamitos, CA, USA, may 2020. IEEE Computer Society.

\end{thebibliography}

\appendix
\section*{Appendix}

\section{Proofs}
\label{app:proof}

\subsection{Proof of Theorem \ref{thm:main} for $\Delta = 0$}
\label{sec:delta=0}
In this section, we prove the security of \newprism Confirmation Rule when $\Delta = 0$.
As a warm-up, we first prove the security of a time-based confirmation for online users in the following lemma.

\begin{lemma}
\label{lem:lem:main_0}
Assume $\Delta = 0$. If an online node is aware of $\tau_\ell$ and confirms $B_\ell(T)$ at time $\tau_\ell + T$ such that $h^0(T) \geq \frac{1}{m} \bar V_\ell(T) + \frac{1}{2} +\delta$, then number of votes received by $B_\ell(T)$ will be greater than number of votes received by any other proposer block at time $t$ for all $t>\tau_\ell + T$, except for probability $e^{-2\delta^2 m}$. Note that $\tau_\ell + T$ is a random time chosen by an honest node to observe the blockchains.
\end{lemma}

\begin{proof}
For $1 \leq i \leq m$, let $H_i(t_1,t_2)$ and $Z_i(t_1,t_2)$ be the number of honest blocks and adversary blocks mined at the $i$-th voter chain in the time interval $(t_1, t_2]$, respectively. Then $H_i(t_1,t_2)$ and $Z_i(t_1,t_2)$ follow Poisson distributions $\texttt{Poi}((1-\beta)\lambda(t_2-t_1))$ and $\texttt{Poi}(\beta\lambda(t_2-t_1))$ respectively.
Let $L_i$ be the maximum lead built by the adversary over the public longest chain on the $i$-th voter chain at time $\tau_\ell$. It have been shown in \cite{sompolinsky2016bitcoin} (Lemma 3) that $L_i$ is first order stochastically dominated by a Geometric random variable $Y \sim \texttt{Geo}(p)$ with $p = \frac{1-2\beta}{1-\beta}$ (i.e., $P(L_i \geq l) \leq P(Y \geq l) = \big( \frac{\beta}{1-\beta} \big)^l$ for $l \in \mathbb{N}$).

Let $U_i(t) = 1$ if the $i$-th voter chain votes for $B_\ell(T)$ at time $\tau_\ell + t$; and 0 otherwise. Then we have $V_\ell(t) = \sum_{i=1}^{m} U_i(t)$.

We prove the theorem by contradiction. Suppose at time $t_1 > \tau_\ell + T$, $V_\ell(t_1)$ is smaller than number of votes received by another proposer block, which implies that $V_\ell(t_1) \leq m/2$.

For $1 \leq i \leq m$, let event $E_i$ be the event that there exists $t_2 \in [\tau_\ell + T, \tau_\ell + t_1]$ such that $L_i + Z_i(\tau_\ell,t_2) \geq H_i(\tau_\ell, t_2)$ on the $i$-th voter chain; let event $F_i$ be the event that there exists $t_2 \in [\tau_\ell + T, +\infty)$ such that $L_i + Z_i(\tau_\ell,t_2) \geq H_i(\tau_\ell, t_2)$ on the $i$-th voter chain. Let event $W_i$ be the event that $i$-th voter chain votes for 
a proposer block at level $\ell$ other than $B_\ell(T)$ at time $T$. Note that $E_i \subset F_i$, and also we have the following important {\bf claim 1}:
\begin{equation*}
   \{U_i(t_1) = 0\} \subseteq E_i \cup W_i, 
\end{equation*}
thus
\begin{equation*}
    U_i(t_1) \geq 1 - \indicator_{E_i\cup W_i} \geq 1 - \indicator_{F_i\cup W_i} \geq 1 - \indicator_{F_i} - \indicator_{W_i}.
\end{equation*}

Sum them up over $1 \leq i \leq m$, we have
\begin{equation*}
    V_\ell(t_1) = \sum_{i=1}^{m} U_i(t) \geq m - \sum_{i=1}^m \indicator_{F_i} - \bar V_\ell(T),
\end{equation*}
then we have
\begin{eqnarray*}
    \sum_{i=1}^m \indicator_{F_i} &\geq& m - \bar V_\ell(T) - V_\ell(t_1) \\
    &\geq& m - \bar V_\ell(T) - m/2 \\
    &=& m/2 - \bar V_\ell(T) \\
    &\geq& m \big (1-h^0(T)\big ) + m\delta.
\end{eqnarray*}

Here we first check the expectation,
\begin{equation*}
    E[\sum_{i=1}^m \indicator_{F_i}] = \sum_{i=1}^m P(F_i) = m P(F_1) = m(1-h^0(T)),
\end{equation*}
by the definition of event $F_i$ and $h^0(t)$. Then we have
\begin{eqnarray}
    \label{eqn:hoeffding_prob}
    &&P(V_\ell(t_1) \leq m/2) \nonumber \\
    &\leq& P(\sum_{i=1}^m \indicator_{F_i} \geq  m(1-h^0(T)) + m\delta)  \nonumber\\
    &=& P(\sum_{i=1}^m \indicator_{F_i} - E[\sum_{i=1}^m \indicator_{F_i}] \geq m\delta)  \nonumber \\
    &\leq& e^{-2\delta^2 m},
\end{eqnarray}
where the last inequality is due to Hoeffding's inequality. (\ref{eqn:hoeffding_prob}) concludes the lemma.

\end{proof}

To complete the proof of Lemma~\ref{lem:lem:main_0}, we now prove {\bf Claim 1}.

\begin{proof}[Proof of Claim 1]
Note that an equivalent expression of event $E_i$ would be the event that there exists time $t_2 \in [\tau_\ell+T,\tau_\ell + t_1]$ such that on the $i$-th voter chain,
\begin{equation*}
    \max_{t_0,t_2:t_0 \leq \tau_\ell, t_2 \in [\tau_\ell+T,\tau_\ell + t_1]}(Z_i(t_0,t_2) - H_i(t_0,t_2)) \geq 0.
\end{equation*}

First, observe that 
\begin{eqnarray*}
&& \{U_i(t_1)=0\} \\
&=&  (\{U_i(t_1)=0\} \cap W_i) \cup (\{U_i(t_1)=0\} \cap W^c_i) \\ 
&\subseteq& W_i \cup (\{U_i(t_1)=0\} \cap W^c_i).
\end{eqnarray*}
Hence, to conclude the proof, it is sufficient to show that
\begin{equation*}
\{U_i(t_1)=0\} \cap W^c_i \subseteq E_i.
\end{equation*}

Second, we make the following observation, hereafter denoted as observation (1):
If the voter chain $i$ does not vote for any proposer block at level $l$ at some time $t>\tau_\ell$, then none of the honest blocks mined over the interval $(\tau_\ell,t]$ (if there are any) are in the voter chain $i$ at time $t$.

Now, define $A_i$ as the event that the voter chain $i$ does not vote for any proposer block at level $l$ at time $T$.
Then, by observation (1), either (i) no honest block was mined over the interval $(\tau_\ell,\tau_\ell+T]$, i.e $H_i(\tau_\ell,\tau_\ell+T)=0$; or (ii) honest blocks were mined over the interval $(\tau_\ell,\tau_\ell+T]$, yet, none of them are in the voter chain $i$ at time $T$.
Since $Z_i(\tau_\ell,\tau_\ell+T)$ is lower bounded by $0$, (i) clearly implies
\begin{equation*}
\max_{t_0,t_2:t_0 \leq \tau_\ell, t_2 \in [\tau_\ell+T,\tau_\ell + t_1]}(Z_i(t_0,t_2) - H_i(t_0,t_2)) \geq 0.
\end{equation*}

\begin{center}
\begin{figure}
     \includegraphics[width=9cm]{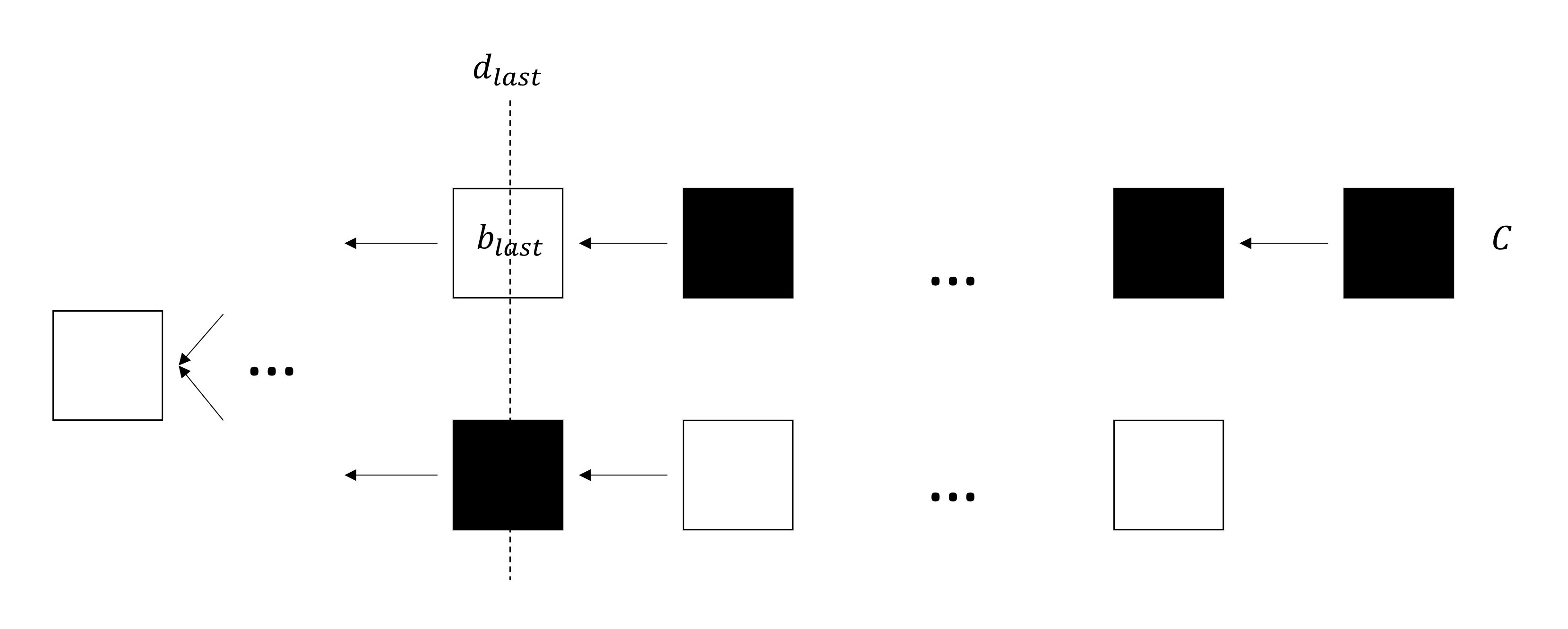}
     \caption{Chain $C$ of adversary blocks. Filled blocks are adversary. Honest blocks mined over the interval $(\tau_\ell,\tau_\ell+T]$ are shown on the same chain below the chain $C$. (In general, they do not have to lie on the same chain.)}
     \label{fig:case_1}
\end{figure}
\end{center}

Next, let's assume that case (ii) happens.
Because every honest block is mined at a different level higher than the previous one, all of the honest blocks mined over the interval $(\tau_\ell,\tau_\ell+T]$ are at a unique level.
Moreover, since none of these honest blocks are in the voter chain $i$ at time $\tau_\ell+T$, there exists a chain $C$ at time $\tau_\ell+T$ that has length larger than or equal to the level of all of these honest blocks and does not contain any of them. See Figure~\ref{fig:case_1} for illustration.
The last honest block, $b_{\mathrm{last}}$, on chain $C$ was mined before time $\tau_\ell$, thus, it is at a level that is smaller than the levels of the honest blocks mined over the interval $(\tau_\ell,\tau_\ell+T]$.
Then, the levels of $C$ that are larger than this last honest block are occupied by adversary blocks.
Let $t_{\mathrm{last}} \leq \tau_\ell$ and $d_{\mathrm{last}}$ respectively be the mining time and level of this last honest block on $C$.
Observe that any honest block mined after time $t_{\mathrm{last}}$ is at a unique level larger than $d_{\mathrm{last}}$ that is matched by a unique adversary block on $C$ that was mined after time $t_{\mathrm{last}}$.
Hence, under case (ii),
\begin{equation*}
Z_i(t_{\mathrm{last}},\tau_\ell+T) - H_i(t_{\mathrm{last}},\tau_\ell+T) \geq 0.
\end{equation*}
Since $t_{\mathrm{last}} \leq \tau_\ell$,
\begin{equation*}
\max_{t_0,t_2:t_0 \leq \tau_\ell, t_2 \in [\tau_\ell+T,\tau_\ell + t_1]}(Z_i(t_0,t_2) - H_i(t_0,t_2)) \geq 0.
\end{equation*}

Finally, together with case (i), we can conclude that $A_i$ implies $E_i$:
\begin{equation*}
A_i \subseteq E_i.
\end{equation*}

Next, we express $W^c_i$ as the disjoint union of the events $A_i$ and $A^c_i \cap W^c_i$.
Observe that $A^c_i \cap W^c_i$ is the event that the voter chain $i$ votes for $B_l(T)$ at time $T$:
\begin{equation*}
A^c_i \cap W^c_i = \{U_i(T)=1\}.
\end{equation*}
Then,
\begin{equation*}
\{U_i(t_1)=0\} \cap A^c_i \cap W^c_i = \{U_i(T)=1\} \cap \{U_i(t_1)=0\}.
\end{equation*}

The event $\{U_i(T)=1\}$ implies the existence of a (possibly adversary) block $b^*$ mined at some level $d^*$ before time $\tau_\ell+T$ that votes for $B_l(T)$ and is in the voter chain $i$ at time $\tau_\ell+T$.
Similarly, $\{U_i(T)=1\} \cap \{U_i(t_1)=0\}$ implies the existence of a time $t_2 \in (\tau_\ell+T,\tau_\ell + t_1]$ such that $b^*$ leaves the voter chain $i$ for the first time after $\tau_\ell+T$ at time $t_2$.
Then, at time $t_2$, there exists a chain $C^*$ that contains $b^*$ and has length $D^*$ and a chain $C$ that does not contain $b^*$ and has length $D \geq D^*$. See Figure~\ref{fig:case_2} for illustration.
Notice that no honest block can be at a level larger than $D^*$ at time $t_2$ because $C^*$ \emph{was} the (public) voter chain $i$ just before time $t_2$.

\begin{figure}
     \includegraphics[width=9cm]{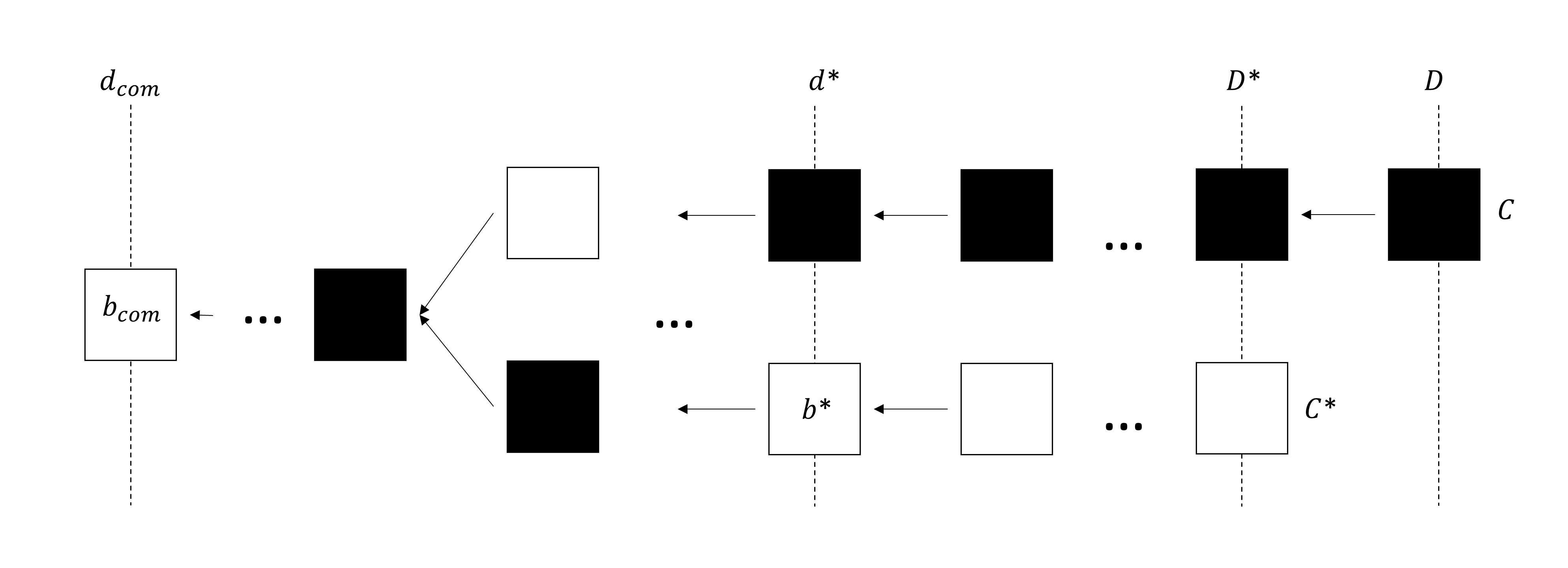}
     \caption{Chains $C^*$ and $C$ with heights $D^*$ and $D$. Filled blocks are adversary.}
     \label{fig:case_2}
\end{figure}

Due to observation (1), none of the honest blocks mined before $b^*$ and after time $\tau_\ell$ are in the prefix of $b^*$ in $C^*$, implying that all of the blocks mined after time $\tau_\ell$ in the prefix $b^*$ are adversary blocks. 
Let $b_{\mathrm{com}}$ with mining time $t_{\mathrm{com}}$ and level $d_{\mathrm{com}}$ denote the last honest common block on the chains $C$ and $C^*$.
Note that $b_{\mathrm{com}}$ lies in the prefix of $b^*$ and is an honest block.
Then, as all of the blocks mined after time $\tau_\ell$ in the prefix $b^*$ are adversary blocks, we know that $t_{\mathrm{com}} \leq \tau_\ell.$ Refer to Figure~\ref{fig:case_2}.

Now, at every level in the interval $(d_{\mathrm{com}},D^*]$, either the chains $C$ and $C^*$ have a common adversary block, or the chains have two distinct blocks, one of which has to be an adversary block.
Hence, every level in the interval $(d_{\mathrm{com}},D^*]$ contains an adversary block that is a descendant of $b_{\mathrm{com}}$.
Then, any honest block mined in the interval $(t_{\mathrm{com}},t_2]$ arrives at a unique level $d$, $d_{\mathrm{com}} < d \leq D^*$, that also contains an adversary block descending from $b_{\mathrm{com}}$.
Since these adversary blocks descending from $b_{\mathrm{com}}$ should have been mined after time $t_{\mathrm{com}}$, i.e in the interval $(t_{\mathrm{com}},t_2]$, we conclude that
\begin{equation*}
Z_i(t_{\mathrm{com}},t_2) - H_i(t_{\mathrm{com}},t_2) \geq 0.
\end{equation*}
Since $t_{\mathrm{com}} \leq \tau_\ell$,
\begin{equation*}
\max_{t_0,t_2:t_0 \leq \tau_\ell, t_2 \in [\tau_\ell+T,\tau_\ell + t_1]}(Z_i(t_0,t_2) - H_i(t_0,t_2)) \geq 0.
\end{equation*}

Finally, we observe that
\begin{equation*}
\{U_i(T)=1\} \cap \{U_i(t_1)=0\} \subseteq E_i,
\end{equation*}
which further implies
\begin{eqnarray*}
&&\{U_i(t_1)=0\} \cap W^c_i \\
&=& (\{U_i(t_1)=0\} \cap A^c_i \cap W^c_i) \cup (\{U_i(t_1)=0\} \cap A_i \cap W^c_i) \\
&\subseteq& (\{U_i(T)=1\} \cap \{U_i(t_1)=0\}) \cup A_i \\
&\subseteq& E_i,
\end{eqnarray*}
which concludes the proof.

\end{proof}

Then we can prove the main theorem. We restate it for $\Delta = 0$ in the following theorem.

\begin{theorem}[Restatement of Theorem \ref{thm:main} for $\Delta = 0$]
\label{thm:main_0}
Assume $\Delta = 0$. If $B_\ell(T)$ is confirmed at time $\tau_\ell + T$,i.e., 
$$h^0(\frac{1}{(1+\delta)m\lambda}\sum_{i=1}^{m} d_i(T)) \geq \frac{1}{m} \bar V_\ell(T) + \frac{1}{2} +\delta,$$ 
then number of votes received by $B_\ell(T)$ will be greater than number of votes received by any other proposer block at time $t$ for all $t>\tau_\ell + T$, except for probability $e^{-2\delta^2 m} + 2e^{-\frac{\delta^2 t_\delta^* \lambda m}{3}}$, where $t_\delta^*$ is the time such that $h^0(t_\delta^*) = \frac{1}{2} + \delta$ for $0 < \delta < \frac{1}{2}$. Note that $\tau_\ell + T$ is a random time chosen by an honest node to observe the blockchains.
\end{theorem}

\begin{remark}
When there is only one single proposer block at level $\ell$ (i.e., $\bar V_\ell(T) = 0$), the confirmation rule is simply confirming the single proposer block when $\sum_{i=1}^{m} d_i(T) \geq (1+\delta)m\lambda t^*_\delta$ by the monotonicity of $h^0(t)$.
\end{remark}

\begin{proof}
We define random variables $H_i$, $Z_i$ and $U_i(t)$, events $E_i$ and $F_i$ ($1\leq i \leq m$) the same as in Lemma~\ref{lem:lem:main_0}.
And again we prove the theorem by contradiction. Suppose at time $t_1 > \tau_\ell + T$, $V_\ell(t_1)$ is smaller than number of votes received by another proposer block, which implies that $V_\ell(t_1) \leq m/2$.
Then similar to the arguments in Lemma \ref{lem:lem:main_0}, we have
\begin{eqnarray*}
    \sum_{i=1}^m \indicator_{F_i} &\geq& m - \bar V_\ell(T) - V_\ell(t_1) \\
    &\geq& m - \bar V_\ell(T) - m/2 \\
    &=& m/2 - \bar V_\ell(T) \\
    &\geq& m \big (1-h^0(\frac{1}{(1+\delta)m\lambda}\sum_{i=1}^{m} d_i(T))\big ) + m\delta.
\end{eqnarray*}
Let event $G$ be the event that $\sum_{i=1}^m d_i(T) \geq (1+\delta) m\lambda T$, i.e., the sum of the depths of the voter blocks is atypically large at time $\tau_\ell + T$. Then we have
\begin{eqnarray}
    \label{eqn:total_prob}
    && P(V_\ell(t_1) \leq m/2)  \nonumber\\
    &\leq& P(\sum_{i=1}^m \indicator_{F_i} \geq  m \big (1-h^0(\frac{1}{(1+\delta)m\lambda}\sum_{i=1}^{m} d_i(T))\big ) + m\delta) \nonumber\\
    &=& P(G,\sum_{i=1}^m \indicator_{F_i} \geq  m \big (1-h^0(\frac{1}{(1+\delta)m\lambda}\sum_{i=1}^{m} d_i(T))\big ) + m\delta) \nonumber\\
    &+& P(G^c,\sum_{i=1}^m \indicator_{F_i} \geq  m \big (1-h^0(\frac{1}{(1+\delta)m\lambda}\sum_{i=1}^{m} d_i(T))\big ) + m\delta) \nonumber\\
    &\leq& P(G) + P(\sum_{i=1}^m \indicator_{F_i} \geq  m \big (1-h^0(T))\big ) + m\delta).
\end{eqnarray}

We first bound the probability $P(G)$. Since we have 
$$h^0(\frac{1}{(1+\delta)m\lambda}\sum_{i=1}^{m} d_i(T)) \geq \frac{1}{m} \bar V_\ell(T) + \frac{1}{2} +\delta \geq \frac{1}{2} +\delta,$$
by the definition of $t^*_\delta$ and the monotonicity of $h^0(t)$, we know $\sum_{i=1}^m d_i(T) \geq (1+\delta)m\lambda t^*_\delta$. And since the depth of a voter block is maximized when no fork happens, we know $\sum_{i=1}^{m} (H_i(\tau_\ell,\tau_\ell + T)+Z_i(\tau_\ell,\tau_\ell + T)) \geq \sum_{i=1}^{m} d_i(T) \geq (1+\delta)m\lambda t^*_\delta$. Let event $\hat G$ be the event that $\sum_{i=1}^{m} (H_i(\tau_\ell,\tau_\ell + T)+Z_i(\tau_\ell,\tau_\ell + T)) \geq (1+\delta) m\lambda T$.
Then we have
\begin{eqnarray}
\label{eqn:first_prob}
    && P(G) \nonumber \\
    &\leq& P(\hat G, T> t^*_\delta) + P(\hat G, T \leq t^*_\delta) \nonumber \\
    &\leq& P(\texttt{Poi}(m\lambda T) \geq (1+\delta) m\lambda T, T> t^*_\delta) + P(T \leq t^*_\delta) \nonumber \\
    &\leq& P(\texttt{Poi}(m\lambda T) \geq (1+\delta) m\lambda T, T> t^*_\delta) \nonumber \\
    &+& P(\texttt{Poi}(m\lambda t^*_\delta) \geq (1+\delta) m\lambda t^*_\delta) \nonumber \\
    &\leq& 2e^{-\frac{\delta^2 t_\delta^* \lambda m}{3}},
\end{eqnarray}
where the last inequality is due to Poisson tail bound.

For the second term in (\ref{eqn:total_prob}), by Lemma \ref{lem:lem:main_0} we have
\begin{equation}
    \label{eqn:second_prob}
     P(\sum_{i=1}^m \indicator_{F_i} \geq  m(1-h^0(T)) + m\delta)  \leq e^{-2\delta^2 m}.
\end{equation}
Combine (\ref{eqn:total_prob}), (\ref{eqn:first_prob}) and (\ref{eqn:second_prob}), we have
\begin{equation*}
    P(V_\ell(t_1) \leq m/2) \leq e^{-2\delta^2 m} + 2e^{-\frac{\delta^2 t_\delta^* \lambda m}{3}},
\end{equation*}
which concludes the proof.
\end{proof}

\subsection{Proof  of Theorem \ref{thm:main} for $\Delta > 0$}
\label{app:proof_sketch}
We outline the steps needed to extend the proof technique in Theorem \ref{thm:main_0} to general cases when $\Delta > 0$.

{\bf \noindent Worst case attack}. For positive $\Delta$, characterizing the worst case attack for the longest chain protocol is challenging, but we are only interested in the regime where the block speed of each voter chain is very slow (i.e. $\lambda \Delta \ll 1$). 
Without loss of generality, we renormalize time such that $\lambda = 1$ to simplify notations.
Recall that $q_t(\lambda_h,\lambda_a,\Delta,\pi)$ is the reversal error probability of a block that has been mined for time $t$ under adversary strategy $\pi$, where $\lambda_h = (1-\beta)$ is the honest mining rate and $\lambda_a = \beta$ is the adversary mining rate. We fix $\beta$ and $t$, and write $f(\Delta, \pi) = q_t(\lambda_h,\lambda_a,\Delta,\pi)$. Let
$$g(\Delta) = \max_{\pi} f(\Delta, \pi),$$
$$\pi^*(\Delta) = \arg \max_{\pi} f(\Delta, \pi).$$
And we know $\pi^*(0)$ is the private attack with pre-mining. For $\Delta > 0$, we have
\begin{eqnarray*}
    &&g(\Delta)  \\
    &=& f(\Delta, \pi^*(\Delta)) \\
    &=& f(0, \pi^*(0)) + \frac{\partial f}{\partial \Delta}(0,\pi^*(0))\cdot \Delta \\
    &~~~~+& \frac{\partial f}{\partial \pi^*}(0,\pi^*(0))\cdot \frac{\partial \pi^*(0)}{\partial \Delta} \cdot \Delta + O(\Delta^2) \\
    &=& f(0, \pi^*(0)) + \frac{\partial f}{\partial \Delta}(0,\pi^*(0))\cdot \Delta + O(\Delta^2) \\
    &=& f(\Delta, \pi^*(0)) + O(\Delta^2).
\end{eqnarray*}
Thus  reintroducing $\lambda$ into the result, we have 
$$q_t(\lambda_h,\lambda_a,\Delta,\pi^*(\lambda,\Delta)) = q_t(\lambda_h,\lambda_a,\Delta, \pi^*(\lambda,0)) + O(\lambda^2\Delta^2),$$ 
i.e., for small $\lambda \Delta$ we can use the error probability under the private attack to approximate the error probability under the worst attack. 

{\bf \noindent Poisson approximation}. Further note that for $\lambda \Delta > 0$, the pure honest chain does not grow as a Poisson process because of forking. \cite{dembo2020everything} showed that the worst case forking and hence the worst case chain growth is achieved when every honest block is delivered with maximum delay $\Delta$. In the worst case forking, let $T_d$ be the time it takes for pure honest chain to reach depth $d$ after reaching depth $d - 1$. Then by the memoryless property of the mining process, $T_d$'s are i.i.d. random variables and we can write $T_d$ as $T_d = \Delta + S_d $, where $S_d$'s are i.i.d.\ and exponentially distributed with rate $\lambda_h$. So we have 
\begin{equation*}
    E[T_d] = \Delta + E[S_d] = \Delta + \frac{1}{\lambda_h}.
\end{equation*}

To further simplify the calculation, we approximate the distribution of $Y_d$ with an exponential distribution with rate $\lambda_h'$, where $\lambda_h' = \frac{\lambda_h}{1+\lambda_h \Delta}$, which is equivalent to model the honest chain growth as Poisson process with rate $\lambda_h'$. 

With the two approximations above, we can simply calculate $h^{\Delta}(t)$ by

\begin{eqnarray*}
    h^\Delta(t) &=& 1 - q_t(\lambda_h,\lambda_a,\Delta,\pi^*(\lambda,\Delta))  \\
    &\approx& 1 - q_t^0(\lambda_h',\lambda_a,).
\end{eqnarray*}

{\bf \noindent Correlation across voter chains}. Note that $h^\Delta(t)$ only captures the behavior of one single voter chain. To complete the proof of the main theorem, we still need some independence across all the voter chains. In the proof of Theorem \ref{thm:main_0}, we implicitly assume that the mining processes or arrival of blocks on different voter chains are independent, which is achieved by the Cryptographic sortition mechanism proposed in Prism. The sortition splits the aggregate mining process, which we model it as a Poisson process when the number of miners is very large, into $m$ independent Poisson processes with equal rates.

Further, following the proof of Lemma 16 in \cite{cryptoeprint:2020:675}, not only the events related to block arrivals are independent, the adversary attacks on different voter chains are also negatively correlated.
That is, $P(\cap F_i) \leq \prod P(F_i)$ conditioned on some events occurring with probability at least $1-\varepsilon$, where $F_i$ is the persistence or liveness failure events on the $i$-th voter chain and $\varepsilon$ is negligible in the security parameter. This result directly implies that the correlations across the voter chains are simply negative for the adversary. Armed with this negative correlation property, the proof of Theorem $\ref{thm:main_0}$ can now be extended to the general non-zero delay case.

\subsection{Proof of Theorem \ref{thm:latency}}
\label{app:latency}

In this section, we discuss the latency of our proposed confirmation rule. We focus on the single proposer block case, i.e., the latency of CR3. Note that although the depth of a voter block is observable for both online and offline nodes, it also greatly depends on the adversary actions. For example, under private attack, each voter chain does not grow with the full rate $\lambda$, hence the confirmation will get delayed compared with the scenario where the adversary follows the protocol. Further, we will see that the adversary can even reduce the depth of the vote by revealing a private chain consist of voter blocks with null votes. Fortunately, we are able to obtain the upper bound of the latency in the single proposer block case by identifying the worst case attack in term of delaying the confirmation.  

{\bf \noindent No attack}. The best case latency of CR3 is achieved when the adversary follows the voting rule and longest chain rule in each voter chain. Under this no attack case, the depth of the vote grows as a Poisson process with rate $\lambda$. Hence the expected latency under no attack is $(1+\delta)m\lambda t^*_\delta/(m\lambda) = (1+\delta) t^*_\delta$.

{\bf \noindent Private attack}. In the private attack, the adversary mines private chains forked from the voter blocks that vote for the single proposer block at level $\ell$, and delays the delivery of every honest voter block with the maximum network delay $\Delta$. We have already seen that the pure honest chain growth under the worst case forking can be approximated with a Poisson process with rate $\lambda_h' = \frac{\lambda_h}{1+\lambda_h \Delta}$. Hence the expected latency under private attack is $(1+\delta)m\lambda t^*_\delta/(m\lambda_h') = (1+\delta) t^*_\delta\lambda/\lambda_h'$.

{\bf \noindent Slow-down attack}.   Next we  show that the worst case latency of CR3 is achieved when the adversary uses the following ``slow down'' attack on all the voter chains.

\begin{defn}[Slow down attack]
\label{def:slow}
On each voter chain, the adversary mines a private chain with null votes forked from the voter block that votes for the single proposer block at level $\ell$. Meanwhile, the adversary delays the delivery of every honest voter block with the maximum network delay $\Delta$. Whenever the length of the private chain is greater than or equal to the public honest chain, the adversary reveals some private blocks to match the length of the public chain and force all honest nodes to mine on the adversary block (we assume tie breaking is in favor of the adversary in the longest chain rule). 
\end{defn}


\begin{lemma}[Optimality of slow down attack]
\label{lem:slow_down}
The slow down attack is the worst case attack in term of reducing the depth of the vote on a voter chain.
\end{lemma}

\begin{proof}
We show that for each sample path, the slow down attack will lead to the smallest depth of the voter block that votes for the single public proposer block at level $\ell$. Since the slow down attack is independent and identical on all the voter chains, we will only focus on one of the voter chains and WLOG let it be voter chain one.

Given a sample path (a sequence of block arrival times $t_1, t_2, \cdots, t_n$ in the time interval $(\tau_\ell + \Delta, \tau_\ell + \Delta + t)$). Let $B_\ell$ be the single public proposer block at level $\ell$, then all honest nodes will receive $B_\ell$ at time $\tau_\ell + \Delta$. By the honest voting rule, the depth of the vote for $B_\ell$ is the depth of the first honest block mined after $\tau_\ell + \Delta$ in the public longest chain, that is also the difference between the length of the public longest chain and the length of the public adversary chain (here the length of a chain only counts blocks mined in the time interval $(\tau_\ell + \Delta, \tau_\ell + \Delta + t)$).

We recall that the minimal length of the public longest chain can be achieved by keeping every adversary block private and delivering every honest block with maximum delay $\Delta$. Although the slow down attack may reveal some private blocks at some time, it never increase the length of the public chain. Therefore, the slow down attack achieves the minimal length of the public longest chain at time $\tau_\ell + \Delta + t$.

On the other hand, the maximum length of the public adversary chain is obtained at the last time when the adversary mines more blocks than the length of the public chain. i.e., the largest $t' \in [0,t]$ such that the number of adversary blocks is no less than the length of the honest chain under the worst case forking in the time interval $(\tau_\ell+\Delta,\tau_\ell+\Delta+t')$ (if the adversary never catches up the honest chain before $\tau_\ell + \Delta + t$, then $t' =0$). Since in the slow down attack the adversary reveals its private blocks whenever it can match the public chain, it also achieves the maximum length of the public adversary chain. Therefore, the slow down attack achieves the minimal depth of the vote for $B_\ell$.
\end{proof}

Let $d_i^{SD}(t)$ be the depth of the vote on the $i$-th voter chain at time $\tau_\ell + \Delta + t$ under slow down attack for $1 \leq i \leq m$. Then $d_1^{SD}(t)$, $d_2^{SD}(t)$, \dots, $d_m^{SD}(t)$ are i.i.d. random variables. Again for simplicity we will approximate the arrival of honest blocks as a Poisson process with rate $\lambda_h' = \frac{\lambda_h}{1+\lambda_h \Delta}$. Let the effective total mining rate be $\lambda' = \lambda_h'+\lambda_a$ and the effective adversary fraction be $\beta' = \lambda_a/\lambda'$. Then the distribution of $D = d_1^{SD}(t)$ is given by the following equations. 

\begin{eqnarray}
    && P(D=0)  \nonumber\\
    &=& P(H_1(\tau_\ell +\Delta,\tau_\ell +\Delta +t) \leq Z_1(\tau_\ell+\Delta,\tau_\ell+\Delta + t))  \nonumber \\
    &=& \sum_{j=0}^{+\infty} e^{-(1-\beta')\lambda' t} \frac{((1-\beta')\lambda' t)^j}{j!} \Big( \sum_{i=j}^{+\infty} e^{-\beta'\lambda' t} \frac{(\beta'\lambda' t)^i}{i!} \Big) \nonumber \\
    &&\label{eqn:dist1}\\
    && P(D=d) \nonumber \\
    &=& \sum_{n=d}^{+\infty} e^{-\lambda' t} \frac{(\lambda' t)^n}{n!} \cdot \Big(\sum_{c = \lceil \frac{n-2d}{2} \rceil_+}^{\lfloor \frac{n-d}{2} \rfloor} \binom{2c}{c} \binom{n-2c}{d} \cdot\nonumber \\
    &&~~~~~\beta'^{n-c-d} (1-\beta')^{c+d} \frac{2d+2c-n}{n-2c}\Big) \nonumber \\
    &&\label{eqn:dist2}
\end{eqnarray}

For $d \geq 1$ in (\ref{eqn:dist2}), conditioned on that $n$ blocks are mined in the interval $(\tau_\ell+\Delta,\tau_\ell+\Delta + t)$, the depth of the vote is equal to $d$ when (i) for the first $2c$ blocks, the number of adversary blocks is equal to the number of honest blocks; (ii) for the remaining $n-2c$ blocks, there are $d$ honest blocks and $n-2c-d$ adversary blocks, and also the number of honest blocks is strictly greater than the number of adversary blocks in any prefix of the sample path (which is a Ballot problem). (i) happens with probability $\binom{2c}{c} \beta'^c (1-\beta')^c$ and (ii) happens with probability $\binom{n-2c}{d} \beta'^{n-2c-d} (1-\beta')^d \frac{2d+2c-n}{n-2c}$. By choosing the proper range for $c$, we get (\ref{eqn:dist2}).

\begin{lemma}[Latency of CR3]
\label{lem:latency_cr0'}
The expected worst case latency of CR3 is $t^w$, where $t^w$ is the solution of the equation $E[d_1^{SD}(t)] = (1+\delta)t_\delta^*\lambda$.
\end{lemma}

\begin{proof}
By Lemma \ref{lem:slow_down}, we have $E[\sum_{i=1}^m d_i(t^w)] \geq E[\sum_{i=1}^{m} d_i^{SD}(t^w)] = m \cdot E[ d_1^{SD}(t^w)] = (1+\delta)t_\delta^* \lambda m$.
\end{proof}

Lemma \ref{lem:latency_cr0'} directly implies Theorem \ref{thm:latency}.

\end{document}